%% file: survey.tex
\documentclass[review]{elsarticle}
\usepackage{color}
\usepackage{graphicx}
\usepackage{fullpage}
\usepackage{multirow}
\usepackage{amsmath}
\usepackage{graphics, epstopdf, epsfig}
\usepackage{multicol}
\usepackage{rotating}
\usepackage{booktabs}
\usepackage[table]{xcolor}
\usepackage{longtable}
\usepackage{rotating}
\usepackage{float}
\usepackage{subfigure}
\usepackage{placeins}
\usepackage{tabu}
\usepackage{lscape}

\usepackage{amsmath}

\usepackage{xcolor}

\journal{Journal of COMCOM}

\begin{document}

\begin{frontmatter}

\title{MAC-Layer Rate Control for 802.11 Networks: Lesson Learned and Looking Forward}


\author{Wei Yin$^{1}$, Peizhao Hu$^{2}$, Jadwiga Indulska$^{1}$, Marius Portmann$^{1}$, Ying Mao$^{3}$ }
\cortext[mycorrespondingauthor]{The second author is the corresponding author. Email: ph@cs.rit.edu}
\address{$^{1}$The University of Queensland, Australia}
\address{$^{2}$Rochester Institute of Technology, USA}
\address{$^{3}$The College of New Jersey, USA}



\input{abstract}

\end{frontmatter}

\input{Introduction}

\input{background}

\input{survey_MAC_layer}

\input{analysis}
\input{environments}

\input{critiques}

\input{open_issue}

\input{conclusion}

\input{acknowledgement}

\section*{Reference}
\bibliographystyle{plain}
\bibliography{survey}

\end{document}

%% file: abstract.tex
\begin{abstract}

Rate control at the MAC-layer is one of the fundamental building blocks in many wireless networks. Over the past two decades around thirty mechanisms have been proposed in the literature. Among them, there are mechanisms that make rate selection decisions based on sophisticated measurements of wireless link quality, and others that are based on straight-forward heuristics. Minstrel, for example, is an elegant mechanism that has been adopted by hundreds of millions of computers, yet, not much  was known  about its performance until recently. The purpose of this  paper is to provide a comprehensive survey and analysis of existing solutions from the two fundamental aspects of rate control --- metrics and algorithms. We also review how these solutions were evaluated and compared against each other. Based on our detailed studies and observations, we share important insights on future development of rate control mechanisms at the MAC-layer. This discussion also takes into account the recent developments in wireless technologies and emerging applications, such as Internet-of-Things, and shows issues that need to be addressed in the design of new rate control mechanisms suitable for these technologies and applications.

\end{abstract}

%% file: Introduction.tex
\section{Introduction}
\label{Introduction and Motivation}
Wireless connectivity is fundamental  in supporting the paradigm shift from computing on desktops to Internet-of-Everything \cite{Cisco-Inc.:2016ab}. 802.11 networks are still one of the most common types of wireless networks, which offer high bandwidth and yet efficient energy consumption \cite{Balasubramanian:2009aa}. In the 802.11 networks, the wireless medium is shared and channel conditions are unpredictable due to node mobility, channel fading and interference \cite{Ngo:2016aa}.  Due to these channel dynamics, wireless link quality may vary over time, which has an impact on the performance of the networks, especially for multi-hop networks. The transmission rate is one of the network parameters that determines how fast a node can send data onto the wireless medium. In principle, if the link quality is good such that signals are strong, higher rates can result in higher goodput and lower the channel occupancy time.  In contrast, if the channel quality is poor,  selecting a higher rate increases the probability of packet drop due to weak signals, hence causing additional retransmissions and wasting channel bandwidth. Instead, a lower rate should be utilised to ensure a higher delivery ratio.

Therefore, adaptively selecting an appropriate transmission rate for the given channel quality becomes an important task in improving the performance of each wireless link. This is the task of rate control at the MAC layer.  


There are two fundamental components at the core of every rate control mechanism: (i) a \emph{metric} used to estimate the performance of different rates under the current channel conditions. For example, throughput and frame loss ratio (FLR) are two commonly used metrics in rate control; the typical objectives when using these metrics are selecting a rate that can maximize the achieved throughput or maintain the FLR below a predefined threshold. (ii) an \emph{algorithm} that defines strategies to quickly select a transmission rate appropriate for the given channel conditions. Prolonging the time spent on an inappropriate transmission rate will degrade the performance.


MAC-layer rate control has been an active research topic for almost two decades. \textbf{Biaz et al.~ \cite{biaz2008rate} and Das et al.~\cite{das2014performance} present a survey of rate control mechanisms designed at the early stage. Rate control mechanisms designed for new technologies, e.g. 802.11n and 802.11ac,  are not covered. They neither summarize the lesson learned from study nor give guidelines for rate control researchers. Moreover, there is no indication of research directions for the next generation rate control mechanisms for technologies such as 802.11aa, 802.11ah, 802.11ax, etc.}

This paper serves as a comprehensive survey of the existing MAC-layer rate control mechanisms, including the state-of-the-art solutions. Drawing from our own experience in working on this research topic, we perform a qualitative and quantitative analysis of these solutions using our classification scheme and analytic criteria. Throughout this survey paper, we share the insight of \emph{``what works''} and \emph{``what does not work''} when designing a new MAC-layer rate control mechanism. We believe that we show  important design principles \textbf{and open research issues} that can be of interest for researchers that work on this research topic. 

The rest of this paper is organized as follows. Section \ref{sec: background} provides an overview of the fundamental mechanics of rate control at the MAC-layer. This is followed by a comprehensive survey of the existing rate control mechanisms developed in the last two decades in Section \ref{survey_on_mac_layer}. Similarities and differences between these existing solutions are drawn to form the analysis and technical discussions in Section \ref{analysis}. The discussions are around the five most significant aspects in rate control mechanisms. Section \ref{environments} discusses techniques that have been used to evaluate these existing solutions. Based on observations from these evaluation experiments, we share important insights on the development of rate control in the MAC-layer in Section \ref{sec: lessonlearned}. Section \ref{sec:newDevelopment} discusses new opportunities and shares our view on the role of rate control for various emerging applications.  Section \ref{sec: conclusion} concludes the paper.

%% file: background.tex
\section{Background of MAC-layer rate control}
\label{sec: background}

Typically, the selection of transmission rate depends on the channel quality --- good channel quality can support higher transmission rates. To accommodate the variation in channel quality, the IEEE standards define a number of discrete transmission rates for each coding method \cite{IEEE80211_standard_2012}. For example, the 802.11b mode has four available rates (1, 2, 5.5 and 11~Mbps), 802.11a and 802.11g supports 8 rates (6, 9, 12, 18, 24, 36, 48, 54~Mbps) and 802.11n supports all above 12 rates.

The goal of MAC-layer rate control in  802.11 networks is to select an appropriate rate that achieves the optimal throughput should the channel conditions change. In this section, we provide an overview on why  MAC-layer rate control is needed and what means are available to perform rate adaptation.

\subsection{Problems concerning rate control}
Data transmission can be impacted by the changes in signal quality and collisions of data or control frames.

\subsubsection{Changes in signal quality}
\begin{figure}[t]
\centering
\includegraphics[width=0.5\linewidth]{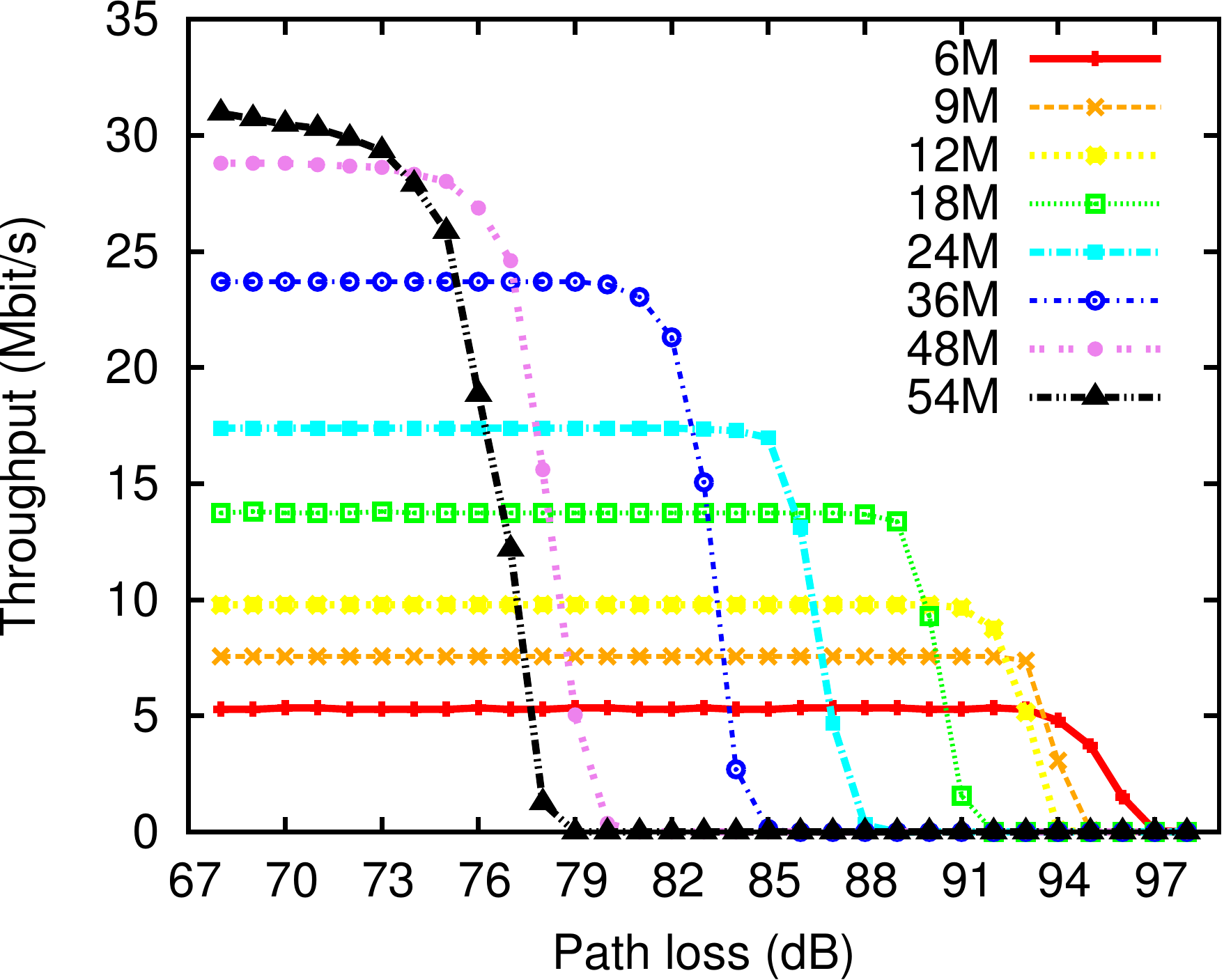}
\caption{Performance of 802.11a fixed rates.}
\label{fig:fixrate_rss}
\end{figure}

In 802.11 networks, the signal quality (or strength) of received frames varies due to a number of reasons, including node mobility or signal fading \cite{mac80211ratecontrol2015}. When two nodes move towards or away from each other, there is a change in signal quality corresponding to the path loss. For example, the free-space path loss, $FSPL=(\frac{4\pi d}{\lambda})^{2}$, describes the power loss as a function of distance $d$ between the transmitter and receiver for a line-of-sight link in an environment that is free of interfering sources; $\lambda$ is the signal wavelength (in meters) \cite{1983aa}. Figure \ref{fig:fixrate_rss} shows the performance of all the 802.11a fixed rates as a function of path loss in free-space \cite{MSWiM2011}. There is an optimal rate that can achieve the optimal throughput for a specific path loss. If the objective is to maximize throughput in the FSPL model, the task of rate control is to determine the optimal rate for a given path loss value. However, the 802.11 networks use the unlicensed spectrums (2.4GHz and 5GHz bands), which are shared with other technologies, such as microwave, Bluetooth and cordless phones. These technologies share the wireless medium and introduce interference in the form of  noise. Substantial increase in  noise can significantly reduce the signal quality as reflected in the signal-to-noise ratio, $SNR=\frac{Signal Power}{Noise Power}$.

\subsubsection{Frame collisions}
While destructive interference will degrade the signal quality in the form of  noise, decodable interfering frames can collide with transmissions and reduce goodput. This interference is typically reflected in the signal-and-interference-to-noise ratio, $SINR=\frac{Signal}{Noise+Interference}$. The signal strength of interfering transmissions is considered separately from noise. Commonly, there are two types of collisions: hidden terminal collision and contention collision. The former occurs when frames from two or more non-carrier-sensing nodes collide \cite{CARA} causing retransmissions. To address this problem, a typical approach is to lower the rate when the throughput drops \cite{minstrel} or frame loss ratio increases \cite{PID}. However this will not address the issue and often makes the matter worse, because lowering the rate will increase the airtime of a frame and therefore increase the probability of collision. RTS/CTS is an effective protocol to address this problem by reserving the channel. In the case of contention collision, frame loss is caused by a large number of nodes competing for channel access \cite{acharya2008}. Under high channel utilization, two or more nodes' backoff timers of the DCF mechanism \cite{IEEE80211_standard_2012} are likely to expire at the same time, thus they attempt to transmit simultaneously causing contention collision. Following the general rule of rate control, the rate will be lowered, making it worse because doing so will increase the airtime, prolonging the channel utilization. The use of RTS/CTS will not address this problem. Techniques, such as clear channel assessment (CCA) or channel state information (CSI), can help to detect the channel activities, but they are not an ideal solution for optimal performance and fairness. In these situations, a scheduling algorithm is needed to coordinate the channel access.


\subsection{Taxonomy of rate control mechanisms}
While rate control mechanisms are categorized by the metrics they  use, they can also be characterized by the following factors:

\subsubsection {Per-frame or adaptation window}
The rapid changes in channel conditions require a responsive adaptation of transmission rates. Ideally, it should select an optimal rate for every frame transmission. However, this requires low-latency hardware that may be dedicated to processing MAC-layer frames. For example, SoftRate \cite{Vutukuru2009} uses Software-Defined radio (SDR) to achieve this per-frame rate selection. \textbf{Some wireless drivers, e.g. Madwifi, on the Linux operating system make a rate selection every \emph{rate adaptation period} (RAP). } RAP is a duration for which an optimal rate will be used for frame transmissions. For every RAP, the performance of one or more rates is calculated based on the desired objectives. The statistics are used to determine the optimal rate for the next RAP. The length of a RAP will have an impact on the responsiveness of the rate control mechanism, and it depends on the implementation of the mechanism.

\subsubsection{``Rate ladder'' or direct rate selection}
The defined discrete rates form a rate ladder for rate adaptation. Given a current rate, $R_{current}$, there are three possible adaptations when channel conditions change: (i) to lower the current rate, $R_{current}^{\downarrow}$, to one level down the ladder, $R_{current}^{-}$, until the base rate; (ii) to increase the current rate $R_{current}^{\uparrow}$, to one level up, $R_{current}^{+}$, until the maximum rate; (iii) to maintain the current rate, therefore, $R_{next}=R_{current}$. Generally speaking, mechanisms can increase or decrease the rate one level at a time. Whereas the direct rate selection approaches compute the performance metric for each rate against a given objective (e.g., maximizing achievable throughput in Minstrel \cite{minstrel}), and select a rate that achieves the optimal utility. It is sensitive to rapid and significant changes in channel conditions.

\subsubsection{Estimation of metrics at sender or receiver}
For most metrics, the computation can be done at the sender. There are metrics that use the MAC-layer acknowledgement (ACK) frame to infer the success or failure of the transmission, while there are others that use the PHY-layer feedback (retry count and rate). In contrast, there are metrics, such as Bit-Error-Rate or SNR, which require measurements of the transmitted frames at the receiver. These measurements are then piggybacked via modified ACK frames to the sender in the rate adaptation processes. There  is a delay in piggybacking the measurements. In addition, modifying the ACK frames introduces overhead and causes compatibility issues.

\subsection{Common features for rate control}
Most rate control mechanisms use the status of frame transmission (Tx) or reception (Rx) of the last frame or adaptation window to make a rate adaptation decision for the next frame or adaptation window. A number of metrics have been considered as the Tx/Rx statuses, including frame transmission time, throughput, signal-to-noise ratio (SNR), and bit-error rate (BER). 

As an example, throughput can be calculated as $\frac{1sec}{T_{tx\_perfect}} * FrameSize$ and the IEEE 802.11 standard \cite{IEEE80211_standard_2012} defines how we can compute the perfect transmission time (transmission time in error free channel),  $T_{tx\_perfect}$: 
\begin{eqnarray}
T_{tx\_perfect} = DIFS + T_{DATA} +   SIFS+ T_{ACK}
\label{eq: Ttx_perfect}
\end{eqnarray}
where $DIFS$ and $SIFS$ are a type of inter-frame spacing defined by the IEEE 802.11 standard; $T_{DATA}$ and $T_{ACK}$ are the transmission time of the data and acknowledgement frames respectively, and they are calculated, as a general form $T_{Frame}$, as
\begin{eqnarray}
T_{Frame} = Ceil((16+8*LEN+6)/NDBPS)\nonumber \\
* T_{SYM} + T_{Preamble} + T_{Signal}
\label{eq: Tframe}
\end{eqnarray}
where $T_{SYM}$ is the symbol interval (defaults to 4~$\mu s$ for 20~MHz channel spacing), $T_{Preamble}$ is the PLCP header preamble duration (defaults to 16~$\mu s$), and $T_{Signal}$ is the duration of the SINGAL BPSK-OFDM symbol (defaults to 4~$\mu s$) as defined in the IEEE 802.11 standard; $LEN$ is the respective frame size, and $NDBPS$ is calculated as $k*r$ where $k$ is a coefficient and $r$ is the bit rate. While the standard describes a reference specification and the fundamental principle is the same, actual implementation depends on individual mechanisms; for example, Minstrel \cite{minstrel}  calculates throughput as $\frac{1sec}{SIFS+T_{DATA}}*FrameSize$, which ignores DIFS and the ACK time. In \cite{Glass:2013aa}, our results showed that the link capacity estimated according to the specification can be far off from the reality. In our study, the abnormal link capacity was caused by a different implementation of the exponential back-off window.

More example metrics are discussed in the specific rate control mechanisms. A collection of other common techniques for rate control is shown in Table \ref{tb: techniques} and briefly described below. 


CSI (Channel State Information) describes various channel properties of a wireless link, including how a signal propagates from the transmitter to the receiver, and the combined effect of fading, power decay and scattering. The sender can adapt transmissions according to the measured CSI values by the receiver and hence improves performance. In a SISO (Single Input Single Output) channel, CSI is only SNR and thus the terminology of CSI is not used. In a MIMO (Multiple Input and Multiple Output) configuration, a CSI matrix is constructed by transmitting signal on each antenna and recording the responses on the receiving antennas. Based on the measured received signals, the node determines the most effective antenna according to the strength of the received signal.

%
%
%

MRR (Multi-Rate-Retry)  is supported in several wireless chipsets including Atheros. In an Atheros chipset, four retry rates (r0, r1, r2 and r3) and corresponding maximum retry counts (c0, c1, c2 and c3) can be specified by the driver and notified to the chipset. r0 can be retried for c0 times if attempts are failed, then r1 will be used, following down to r3 until success. The rate control mechanism usually decides the setting of retry rates and maximum retry counts. MRR is used to deal with short term channel variations, while rate control is for long-term channel variations.

Rate adaptation Period (RAP), as described earlier, defines a period of how long a rate will be utilised for transmitting once it is selected as the best transmission rate. In many common systems, rate selection is based on RAP, which does not require low-latency hardware, such as FPGA, that is capable of selecting rate on a per-frame basis.

Lookaround rate is a sampling approach used by some rate control mechanisms including Minstrel. A rate is used for a whole rate adaptation period once it is selected. The problem for this method is that the mechanism cannot gather fresh statistics about other rates than the selected rate. Therefore, it may make wrong rate decisions with outdated statistics. Minstrel uses a small percentage of frames to try Lookaround rates (rates other than the selected best rate) to probe performance. It sacrifices throughput for up-to-date statistics about Lookaround rates, because a Lookaround rate usually performs worse than the best selected rate.

Frame aggregation is a feature used in a MIMO environment. As the MIMO technology significantly increases the transmission rate, the time to transmit a frame reduces sharply. Assuming that the channel coherence time is constant, we can transmit more data using a high MIMO rate compared to a lower SISO rate during the channel coherence time. Hence,  bundling multiple frames in a single frame is proposed  to minimize the communication overhead, e.g., backoff, interframe spacing, frame header and acknowledgement.

\begin{table*}
\centering
\caption{Common techniques for rate control}
\label{tb: techniques}
{\begin{tabular}{l*{4}c}\toprule
Technique & \multicolumn{4}{p{0.7\linewidth}}{Description}  \\\bottomrule

Channel State Information (CSI)  &\multicolumn{4}{p{0.7\linewidth}}{Describe various channel properties of a wireless link} \\

Multi-Rate-Retry (MRR)  &\multicolumn{4}{p{0.7\linewidth}}{Define the retry policies for failed frames in the driver; when a frame fails to transmit, it will retransmit using the specified retry rates} \\

Rate Adaptation Period (RAP) &\multicolumn{4}{p{0.7\linewidth}}{Define how long a rate will be utilized for transmiting frames once it is selected as the transmission rate.} \\

Lookaround rate  &\multicolumn{4}{p{0.7\linewidth}}{Transmit small portion of frames using other (typically higher) rates rather than the current rate to probe the performance of these rates} \\

Frame aggregation &\multicolumn{4}{p{0.7\linewidth}}{Bundle multiple frame in a single frame in order to minimize the communication overhead, e.g., backoff, interframe spacing, frame header and acknowledgement.}

\\\bottomrule
\end{tabular}}
\end{table*}

%% file: survey_MAC_layer.tex
\section{Existing Mechanisms for Rate Control}
\label{survey_on_mac_layer}

\begin{figure*}[t]
\centering
\includegraphics[width=\textwidth]{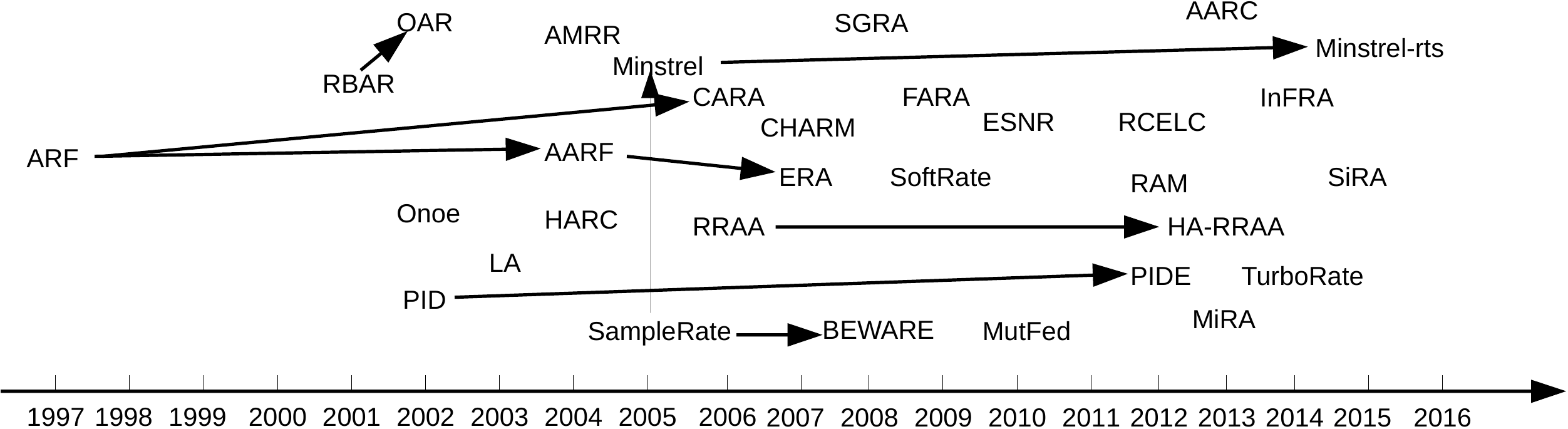}
\caption{A family tree of MAC-layer rate control mechanisms. Each arrow indicates an inheritance relationship.}
\label{fig:history}
\end{figure*}

MAC-layer rate control has been an active research area for almost two decades. In this section, we present a survey of the existing MAC-layer rate control mechanisms for  802.11 networks. Fig. \ref{fig:history} shows a family tree (sorted by the timeline) of these mechanisms. In the figure, arrows indicate the inheritance relationship between two connected mechanisms. In the discussion, we group the mechanisms according to the \emph{metrics} that they use to evaluate the channel or link quality. Through our literature review, we have identified metrics based on (i) consecutive transmission result, (ii) frame loss ratio, (iii) transmission time, (iv) throughput, (v) signal to noise ratio, (vi) bit error rate, (vii) frame error rate, and (viii) combined metrics. 

\subsection{Based on consecutive transmission result (CTR)}
The earliest rate control mechanisms select rates based on the results of previous transmissions, usually on the number of successive transmissions. 

\subsubsection{Automatic Rate Fallback (ARF)}

\begin{figure}[t]
\centering
\includegraphics[width=0.48\textwidth, angle=0]{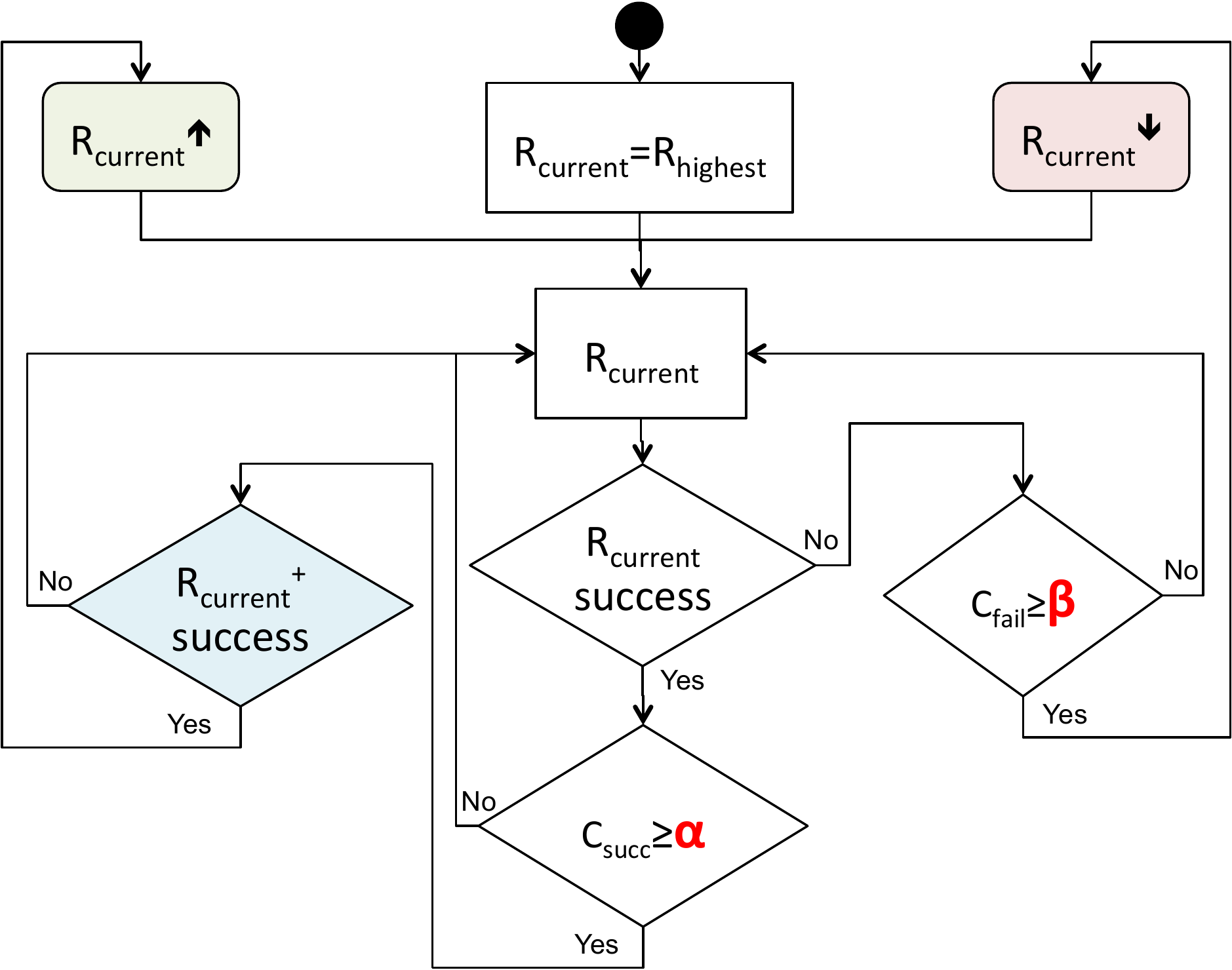}
\caption{Operations of ARF.}
\label{fig:ARF}
\end{figure}

ARF \cite{Karmerman1997} gradually increases or decreases the rate based on the results of previous successive transmissions. It maintains two counters, the number of successes $C_{succ}$ and failures $C_{fail}$ depending on whether acknowledgement frames are received. Fig. \ref{fig:ARF} illustrates the operations of ARF. Starting with the highest rate, as the current rate $R_{current}$, it checks the transmission result of $R_{current}$. If success, it increases $C_{succ}$ and checks whether $C_{succ}\ge\alpha$, where $\alpha$ is a rate increase threshold (defaults to 10 in ARF). It will maintain the current rate, if $C_{succ}\le\alpha$. After ten consecutive successes, ARF will probe the performance of a higher rate $R_{current}^{+}$ that is one level above the current rate. If this probing frame (or \emph{trial frame}) succeeds, it will increase the rate from $R_{current}$ to $R_{current}^{+}$; otherwise, it falls back to the current rate. In the cases when the current rate has failed, ARF increases the $C_{fail}$ and checks $C_{fail}\ge\beta$; where $\beta$ is the rate drop threshold  (defaults to 2). When the failure count reaches the threshold, it decreases from the current rate, $R_{current}$, until the base rate. Otherwise, it falls back to the current rate. Because ARF uses the two counters to accumulate the successive transmission results, a success transmission will reset $C_{fail}$ and a failure transmission will reset $C_{succ}$.

One advantage of ARF is its simplicity. It requires only a timer and two counters to keep track of previous transmission results. ARF generally works well in extremely good or bad channel conditions. However, it is not an efficient mechanism. Because ARF ignores the performance of the current rate, it will attempt to increase the rate after ten consecutive successful transmissions even though the current rate is the optimal rate. It will not converge at an optimal rate. In addition, ARF gradually increases or decreases rate rather than directly jumping to the best rate. This algorithm is unresponsive to rapid, yet significant, changes in channel conditions.

\subsubsection{Adaptive ARF (AARF)}
Lacage et. al. \cite{Lacage2004a} proposed AARF aiming to improve the performance of ARF. The goal of AARF is to converge at the optimal rate by dynamically adjusting the threshold $\alpha$ in ARF. The idea is to double the value of $\alpha$ whenever a trail frame has failed using the next higher rate $R_{current}^{+}$. This change  reduces the number of failed probing frames (overhead) at the higher rate. The threshold $\alpha$ is reset to the initial value 10 when the rate is decreased. 

It slows the response of the mechanism when the rate  needs to be increased, due to the doubling of the $\alpha$ value. In addition, both ARF and AARF will increase the rate to the next level, if the transmission of the first trial frame at the next higher rate $R_{current}^{+}$ is successful after $\alpha$ successful transmissions. In \cite{Wong2006}, Wong et. al. show that the probability of success for a single trial frame is usually higher than 50\%. Increasing the rate solely based on this criterion can potentially lead to incorrect rate adaptation decision, causing frame loss.



\subsubsection{Collision-Aware Rate Adaptation (CARA)}
In wireless networks, frame losses may be caused by collisions or channel errors. CARA  \cite{CARA} improves ARF by incorporating two techniques to differentiate collision frame losses from channel error losses. 

The mandatory technique is called \emph{RTS probing}. If a transmission fails, RTS/CTS will be activated for the first retransmission because the failure may be due to hidden terminal collisions. If this retransmission fails as well, the failure is assumed to be caused by channel errors because the RTS/CTS exchange has already reserved the channel, hence a lower rate should be used for the next retransmission. This mechanism works effectively to avoid occasional frame collisions. But in the situations where collisions are more than an occasional event, this solution will suffer a failure of every first transmission attempt. An optional technique is called \emph{CCA detection}, which serves as a supplement to RTS probing. CARA uses the \emph{busy} counter to determine the channel occupancy during a frame transmission. The CCA technique is used to support the backoff procedures of DCF \cite{CARA}; for example, if CCA is busy, a node freezes its backoff process and resumes when the channel becomes idle. CCA is used to detect collision during the SIFS time, after a node transmitted a data frame and expects an acknowledgement. CARA will conclude the occurrence of a collision if the CCA is assessed as busy during this time, while the expected ACK reception does not start. As a result, the node will retransmit the data frame without increasing $C_{fail}$ and lowering the transmission rate. 

\subsubsection{Effective Rate Adaptation (ERA)}
The rate adaptation in ERA \cite{shaoentechrep2007} is similar to AARF, but it offers a strategy to differentiate losses due to hidden terminal collisions and channel errors. This strategy does not incorporate the RTS/CTS mechanism as CARA does, hence can provide better throughput. If a frame is lost at rate r, the frame is fragmented into two fragments, one very short and the remainder. The 802.11 standard specifies that the channel is reserved for all remaining fragments once the first fragment gets access to the channel. Since the first fragment is very short, it is less likely to experience collision. If it is successful, the mechanism infers that the loss is due to collision, otherwise it is due to channel errors and the base rate is utilised for retransmission. If collision is inferred, the rate will not be decreased. The mechanism is based on an assumption that the short fragment will never experience collisions. This may lead the mechanism to a wrong action.

\subsection{Based on frame loss ratio (FLR)}
Some mechanisms adjust the rate with a goal to maintain the frame loss ratio, $FLR=1-\frac{frames\_received}{frames\_sent}$,  to be within a predefined range.

\subsubsection{Robust Rate Adaptation Algorithm (RRAA)}
Unlike ARF and AARF, RRAA \cite{Wong2006} adapts rate based on the FLR calculated over a time window (e.g., RAP), instead of relying on the success or failure of a single trial frame at the higher rate. In RRAA, a rate is selected for the next time window rather than the next frame. For each supported rate, RRAA defines two thresholds: \emph{Maximum Tolerable Loss threshold (MTL)} and \emph{Opportunistic Rate Increase threshold (ORI)}. The MTL and ORI set the upper and lower bound for FLR. With these two thresholds, we adjust the next rate $R_{next}$ as follow.
\[
    R_{next}= 
\begin{cases}
    R_{current}^{-}, & \text{if } FLR \ge MTL\\
    R_{current}^{+}, & \text{if } FLR \le ORI \\
    R_{current},        & \text{otherwise}
\end{cases}
\]

The value of MTL and ORI is determined by $\alpha L(R)$ and $\gamma L(R)$ respectively; where $\alpha$ and $\gamma$ are tunable parameters, and the \emph{critical loss ratio} $L(R)$ is calculated as
\begin{eqnarray}
L(R) = 1 - \frac{TP(R_{current})} {TP(R_{current}^{-})}
\label{eq: lossratio}
\end{eqnarray}
where $TP(x)$ is the throughput of a rate under perfect channel conditions, and is calculated as $\frac{1sec}{T_{frame}}*FrameSize$; $T_{frame}$ is the perfect transmission time of a frame, defined in Eqn. (\ref{eq: Tframe}). When calculating the \emph{critical loss ratio} $L(R)$, RRAA assumes that the next lower rate has a frame loss ratio of zero, which overestimates the link quality. 

Similar to the RTS probing in CARA, RRAA has a technique called \emph{adaptive RTS} (A-RTS) to differentiate collision losses from channel error losses. In the A-RTS process, RRAA defines a window (\emph{RTSwnd}, and it is initialized to zero to disable RTS/CTS at the beginning) in which RTS frame will be used to avoid collisions. The value of \emph{RTSwnd} depends on whether (i) the frame is lost without RTS frame, (ii) the frame is lost with RTS frame or transmission  succeeds without RTS frame, and (iii) transmission succeeds with RTS frame.
\[
    RTSwnd= 
\begin{cases}
    RTSwnd+1, & \text{if } case\ (i)\\
    half(RTSwnd), & \text{if } case (ii) \\
    unchanged,        & \text{otherwise}
\end{cases}
\]

Within each $RTSwnd$ time interval, a counter is used to keep track of the remaining number of frames to be transmitted with RTS frame. The A-RTS process decreases the counter by one each time a frame is sent out using RTS/CTS, and is terminated when this counter reaches zero. Note that the use of RTS/CTS will introduce overhead, similar to other RTS/CTS approaches.

\subsubsection{History-Aware RRAA (HA-RRAA)}
Pefkianakis et al. \cite{Ioannis2013} proposed improvements for the following three problems in RRAA. The first problem is rate oscillation when RRAA fails to converge at a rate that helps to achieve optimal throughput. RRAA increases the rate when FLR drops below a threshold. This causes an increase in FLR. Subsequently, RRAA reduces the rate to compensate the increase of FLR. Due to low FLR, RRAA again increases the rate and creates the oscillation. HA-RRAA addresses this problem by dynamically adjusting an adaptive window, which specifies the duration for which the current rate should be used. The adaptive window is similar to the rate adaptation window in other mechanisms, but it is not static. The adaptive window $W_{adaptive}$ is computed as
\begin{eqnarray}
W_{adaptive} = W * 2 ^{n} * max(1,\frac{FLR}{10\%})
\label{eq: harraa_tr}
\end{eqnarray}
where $W$ is the channel coherence time that characterizes channel stability \cite{Ioannis2013}, $1\leq n \leq 10$ is the number of consecutive failures in attempting to increase rate, and $FLR$ is the frame loss ratio for the current rate. Essentially, the length of $W_{adaptive}$ increases proportionally to the increase of FLR over 10\% and is doubled for every failed attempt to increase rate. This enhancement extends the duration a rate is used and reduces the occurrence of rate oscillation. The $W_{adaptive}$ will be reset if the rate increase is successful or the rate decreases.



For  802.11a networks, HA-RRAA predefines the number of frames to be transmitted for each rate, ranging from 6-40 frames. Normally, higher rates should be able to transmit more frames successfully, but this is not true in deteriorating link conditions. To address frame loss, HA-RRAA employs a monitoring process. Let $n$ be the predefined frame number for a rate. HA-RRAA will monitor the performance in frame loss every $min(n, 10)$ frames. It will abort using the selected rate if the monitoring results show poor performance.

Finally, HA-RRAA takes into account the RTS/CTS overhead before activating this collision avoidance technique. By comparing the transmission times for data and acknowledgement frames, $T_{DATA+ACK}$, to the transmission times for RTS and CTS frames, $T_{RTS+CTS}$, HA-RRAA engages the RTS/CTS mechanism if $T_{DATA+ACK} \ge k*T_{RTS+CTS}$, where $k$ is a coefficient fixed to 1.5 in the prototype. 



\subsubsection{Onoe}
\begin{figure}[t]
\centering
\includegraphics[width=0.4\linewidth]{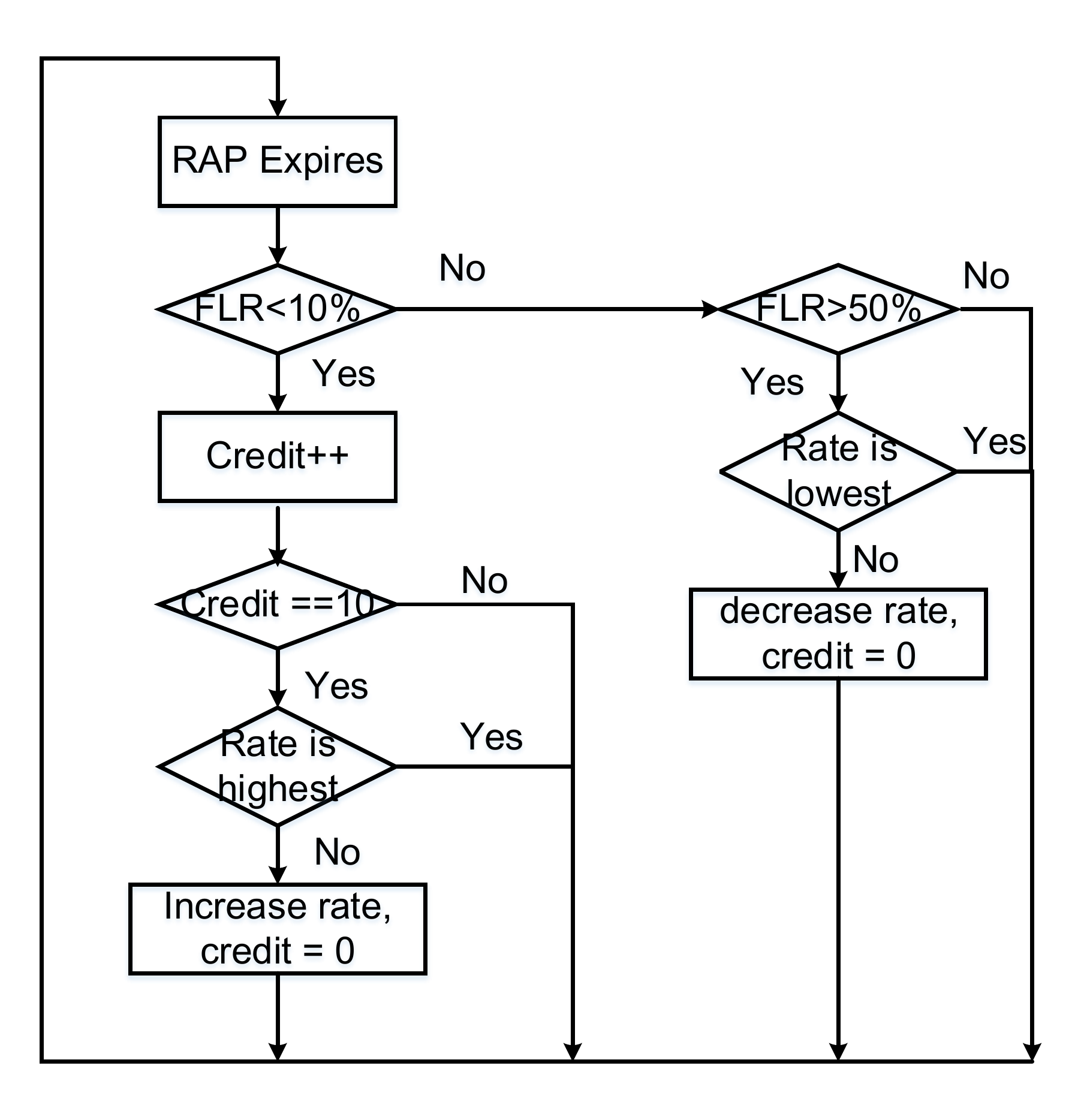}
\caption{How Onoe works.}
\label{fig:Onoe}
\end{figure}

Onoe \cite{Onoe}, as shown in Fig. \ref{fig:Onoe}, makes an indirect use of the FLR. Rather than making rate adaptation decisions based on FLR, it introduces an accumulative credit based scheme, which is changed depending on the averaged number of frame retries in a one-second window.

By default, Onoe sets the initial rate to 24 Mbit/s for both IEEE 802.11g and IEEE 802.11a while it uses 11 Mbit/s for IEEE 802.11b. The credit for the initial rate is initialized to 0. The mechanism increases the number of credits when less than 10\% of frames, in a one-second window, need a retry, while decreasing it otherwise. When the current rate achieves 10 or more credits, Onoe increases the rate to the next higher level, whereas it decreases the rate to the next lower level if 10 or more frames have been sent and the average of retries per frame was greater than one.
Onoe is relatively conservative, because it does not increase the current transmission rate when it detects good quality channel opportunities, but waits until the credit value reaches the threshold. As a consequence, Onoe misses some opportunities of performance enhancement. However, this enables Onoe to be stable. Experiments in \cite{Bicket2005} show that Onoe performs better than ARF and AARF on links where the highest rate experiences significant losses but still has the best throughput. This is because ARF and AARF will decrease the transmission rate when two consecutive failures occur while Onoe can tolerate 50\% frame loss ratio, because it only reduces the rate when the average of retries per frame was greater than one.

\subsubsection{AMRR}

AMRR \cite{Lacage2004a} is proposed for high latency systems. Here, latency refers to the communication delay between the rate control mechanism at the MAC-layer and the wireless interface at the Physical layer. AMRR introduces a new feature called the \emph{multi-rate retry} (MRR) chain, which proposes four candidate rates (i.e., $r0$, $r1$, $r2$ and $r3$) to attempt in case retransmissions are necessary. It shares similar characteristics as other MRR techniques described in Section \ref{sec: background}. For example, the algorithm will first attempt to use rate $r0$ for transmission; if the transmission fails $c0$ times, then rate $r1$ is attempted for retransmission; in the case where rate $r1$ also fails $c1$ times, then rate $r2$ and $r3$ are attempted in the same manner until the base rate is reached. In AMRR, $r0$ is the best rate. $r1$ and $r2$ are subsequent lower rates of $r0$. $r3$ is always set to the base rate.


If more than 10 frames have been transmitted in the last period and the FLR is less than 10\%, AMRR increases $r_{0}$ to the next higher level. The retry chain fails if the FLR is more than 33\% during the last period and thus the transmission rate $r_{0}$ is decreased to the next lower level, until the base rate. AMRR can handle the short-term channel variation. The long-term channel variation is handled by a binary exponential backoff mechanism which adapts the length of the period to change the value of the four rate-count pairs.

\subsubsection{PID}
PID \cite{PID} adapts rates based on a \emph{proportional-integral-derivative} (PID) controller. In essence, the controller is a control loop feedback mechanism that tries to minimise the difference (i.e., \emph{error} in the control system's term) of the current and target frame loss ratio (i.e., $FLR_{current}$ and $FLR_{target}$ respectively) as a result of switching to a new transmission rate. By default, the $FLR_{target}$ is set to 14\% for all rates. 

To determine the appropriate transmission rate, the controller computes an adjustment value, $adj$, as follow.
{\footnotesize 
\begin{eqnarray}
adj =  \gamma * (1+sharpening) * (error_{current} - error_{last}) \nonumber \\ 
    +  \alpha * error_{current} + \beta * error_{avg}
\label{eq: pid}
\end{eqnarray}
}
where $error_{current}$ is the current error, and its value is calculated as $FLR_{target}-FLR_{current}$.
$error_{avg}$ is the average of recent errors, while $error_{last}$ is the last error. In addition, there are four tuneable parameters.  $sharpening$ is a smoothing factor (non-zero when fast response is needed), whereas $\alpha$, $\beta$ and $\gamma$ are the corresponding \emph{proportional}, \emph{integral} and \emph{derivative} coefficients. 

Using Eqn. \ref{eq: pid} the mechanism computes the adjustment value, $adj$, at the end of each rate adaptation period and decides on whether to switch to a new transmission rate. When $adj$ is positive, the new rate $R_{new}$ is set to the highest rate, in the range of $R_{current} \leq R_{new} \leq (R_{current}+adj)$, and its error (i.e., the difference between the target and respective frame loss ratio) is no more than the error of the current rate, $R_{current}$. When $adj$ is negative, the new rate $R_{new}$ is set to the lowest rate, in the range of $(R_{current}+adj)\leq R_{new}\leq R_{current}$, and its error is no more than the error of the current rate. No rate adaptation is required, if $adj$ is equal to zero.

\subsubsection{PIDE}

The idea of PID is based on the control system's principle of a feedback loop. That is, defining a target frame loss ratio (FLR) --- e.g., the target FLR in PID is fixed at 14\% --- let the rate control mechanism to adapt its rate to meet that target by minimizing the difference between the current FLR and the target FLR. It is very straight forward; such that PID drops the sending rate when FLR is greater than 14\%, while increases the sending rate when FLR is less than the target FLR. The most significant problem with PID is its instability in rate selection, as discovered in \cite{MSWiM2011}.

During the adaptation process PID will increase the rate whenever the current FLR is below the target threshold, regardless of the current channel conditions which may not be sufficient to support the higher rate. The consequence of this is the FLR of the higher rate is higher than the target threshold and this causes PID to drop the rate again. Hence, this results in oscillation in rate selection.

Wei et al. \cite{MSWiM2011} implement a verification mechanism in PIDE, which probes the achieved throughput of the proposed rate compared to the current throughput. If the proposed rate achieves higher throughput than the current sending rate, the algorithm  should select the proposed rate. Otherwise, the rate adaptation requests should be ignored. The same rule applies for requests either to increase rate or to decrease rate.

The mechanism uses information passed by the PID controller to detect requests on rate adaptation at the end of each rate adaptation period. When a new rate is proposed (at the end of the last adaptation period) and the proposed rate has not been verified, PIDE then sends $n$ frames (three frames by default) using the proposed rate in the current rate adaptation period. Other frames within the adaptation period will continue to use the current rate. Each frame is associated with a status that records a number of statistics regarding the transmission (e.g., retry count). PIDE collects these statistics and maintains a table of performance information for each rate. At the end of the current adaptation period, PIDE compares the achieved throughput to decide whether to use the proposed rate in the next rate adaptation period. The achieved throughput $TP$ is calculated as
\begin{eqnarray}
TP = (1-FLR) * (1s~/T) * FrameSize
\label{eq: linkcapacity}
\end{eqnarray}
where FLR is the frame loss ratio, 1$s$ is one second, and $T$ is calculated as $DIFS+T_{DATA}+SIFS+T_{ACK}$ \cite{IEEE80211_standard_2012}; and $T_{DATA}$ and $T_{ACK}$ are the respective transmission times of the DATA and ACK frames.

\subsection{Based on transmission time}
As described in Section \ref{sec: background}, the IEEE 802.11 standard specifies the transmission time in perfect, in other words \emph{error-free}, channel conditions. Typically, the transmission time includes all the PHY and MAC overheads such as PHY preamble, SIFS, DIFS, slot time, ACK, and random backoff.

\subsubsection{SampleRate}

SampleRate \cite{Bicket2005} periodically sends a number of data frames as \emph{sample} frames at a rate other than the current rate to gather statistics and to estimate whether another rate would provide better performance.

For each transmission rate, SampleRate calculates the \emph{average transmission time} (ATT) every 10 seconds based on the transmission results, including frame retries. The estimated transmission time is computed according to Eqn. (\ref{eq: Ttx_perfect}). The rate which has the smallest ATT is selected for use in the next control period. 


Every tenth frame is treated as a sample frame to be transmitted at another rate, which is randomly selected from a set of available rates. Otherwise, it sends packets at the rate which has the lowest ATT. Experiments in \cite{Bicket2005} show that SampleRate performs similar to, some time better than, the best of the ARF, AARF and Onoe mechanisms even on lossy links.

\subsubsection{BEWARE}
Wang et al. \cite {BEWARE} proposed another rate control mechanism --- BEWARE, which selects an optimal rate that has the lowest average transmission time. The authors argued that background network traffics should be considered when selecting a rate. SampleRate estimates transmission time using Eqn. (\ref{eq: Ttx_perfect}), which does not consider the time that background traffic causes the CSMA/CA to freeze its counting down. BEWARE introduces mechanisms to measure this time and adds it to the total transmission time of a rate. By considering background traffic, BEWARE becomes load-aware.

\subsection{Based on throughput}
From transmission time, we can compute the achievable throughput over one second as $\frac{1sec}{T_{tx\_perfect}}*FrameSize$, where $T_{tx\_perfect}$ is calculated as Eqn. (\ref{eq: Ttx_perfect}).

\subsubsection{Minstrel}

In Minstrel  \cite{minstrel}, the maximum achievable throughput ($TP$) of each rate is calculated periodically and the rate that is capable to achieve the maximum throughput is selected. The throughput for the given channel conditions is calculated as
\begin{eqnarray}
TP = P_{new} * (\frac{1sec}{T_{tx\_perfect}}) * FrameSize
\label{eq: minstrel_Thp}
\end{eqnarray}
where $P_{new}$ is the weighted probability of success for the current rate adaptation (RAP) window that is to be used by the rate selection process and is computed in Eqn. (\ref{eq: Pewma}), and $T_{tx\_perfect}$ is the time required to successfully deliver a frame in perfect channel conditions; therefore, $\frac{1sec}{T_{tx\_perfect}}$ is the number of packets that can be transmitted in one second window under the perfect channel conditions.

For every interval of 100~ms (default RAP length), Minstrel measures the statistics of frame delivery and computes $P_{new}$ using the Exponential Weighted Moving Average (EWMA), which controls the balance of influence of both old and new packet delivery statistics. 
\begin{eqnarray}
P_{new} = (1 - \alpha) *  P_{this\_interval}+ \alpha * P_{previous}
\label{eq: Pewma}
\end{eqnarray}
where $P_{this\_interval}$  represents the probability of success for the interval before rate selection and and is computed as the ratio of the number of successful transmissions against the number of attempts. $P_{previous}$ represents the moving average probability of success for the last interval. By default the smoothing factor $\alpha$ is set to 75\%, which means historical throughput measurements have more weight on new rate selection.

According to the mac80211 framework source code within the 2.6.35 Linux-wireless kernel, Minstrel calculates the perfect transmission time, $T_{tx\_perfect}$, as 
\begin{eqnarray}
T_{tx\_perfect} = SIFS+T_{Frame}
\label{eq: Minstrel Ttx}
\end{eqnarray}
where $SIFS$ is a type of inter-frame spacing defined by the IEEE 802.11 standard \cite{IEEE80211_standard_2012}; $T_{Frame}$ is the transmission time of the data frame, and it is calculated as Eqn. (\ref{eq: Tframe}). For frame size, Minstrel uses a fixed size of 1200 bytes. We can derive from the Eqns. (\ref{eq: Tframe}) and (\ref{eq: Minstrel Ttx}) that Minstrel's throughput estimation  depends on two parameters, the probability of success to deliver a frame and the bit rate it uses. 

To account for retransmissions Minstrel uses the Multi-Rate Retry (MRR) chain, which proposes four candidate rates (i.e., $r0$, $r1$, $r2$ and $r3$) to attempt in case re-transmissions are necessary (its details are described in Section \ref{sec: background}).

\begin{table}[th]
\center
\caption{Retry preferences}
\begin{tabular}{|c|c|c|c|}
\hline
Attempt & \multicolumn{2}{|c|}{Lookaround rate}  & Normal \\
 	& $RR < BTR$ & $RR > BTR$ &  rate\\
\hline
$r0$ &BTR& RR & BTR\\
$r1$ &RR& BTR & NBTR\\
$r2$ &BPR& BPR & BPR\\
$r3$ &BR& BR & BR\\
\hline
\end{tabular}
\label{Table:retryPreferences}
\end{table}

To determine the optimal rate for a given channel condition, Minstrel dedicates 10\% of its traffic to probe the performance statistics of other rates by randomly selecting a rate (as \emph{Lookaround rate}) that is not currently in use. For this 10\% of data traffic, as shown in Table \ref{Table:retryPreferences}, the retry preferences are the best throughput rate (BTR), the random rate (RR), the best probability rate (BPR) and the base rate (BR) if the randomly selected rate ($RR$) is lower than the current best throughput rate ($BTR$); otherwise they are the random rate, the best throughput rate, the best probability rate, and the base rate. For the other 90\% of traffic (\emph{normal packets}), the retry preferences are the best throughput rate, next best throughput rate (NBTR), the best probability rate and the base rate.

As analysed in \cite{mac80211ratecontrol2015}, in a  collision environment, FLR based rate control mechanisms decrease the transmission rate to the lowest rate, which makes collisions even worse. Minstrel is robust in the collision environment, because it does not decrease the rate. This is an advantage of throughput based mechanism over the CTR or FLR based mechanisms.

\subsubsection{Minstrel-rts}
Although Minstrel is robust in a collision environment, hidden terminal collisions cause frame losses, which decrease the network performance. RTS/CTS is generally recognised as a good approach to mitigate hidden terminal collisions. However, it cannot be activated all the time due to the significant increase in overhead. To address the problem of hidden terminals and at the same time account for the overhead, Minstrel-rts \cite{mac80211ratecontrol2015} adaptively activates the RTS/CTS mechanism depending on whether using the mechanism achieves higher throughput. For \textit{normal} data frames, Minstrel-rts turns off RTS/CTS for the best rate (i.e., $r_{0}$ in MRR). In a collision free environment, the best rate is very likely to be successful so RTS/CTS should be turned off for it. RTS/CTS is randomly turned  on with a probability of 10\% for all retry rates in the MRR. Retry rates are only used when the best rate fails. Frame loss may be due to hidden terminal collisions. Therefore Minstrel-rts turns it on with a small probability to probe the performance statistics for having RTS/CTS turned on; so that the hidden terminal problem can be detected earlier. For \textit{lookaround} frames, RTS/CTS is not used for the random rate (higher than the current rate), because it will only be successful when link quality increases. If this random rate fails the frame will be sent with the best rate without RTS/CTS. The other retry rates use the RTS/CTS mechanism similarly to normal frames.

\begin{figure}[t]
\centering
\includegraphics[width=0.5\textwidth]{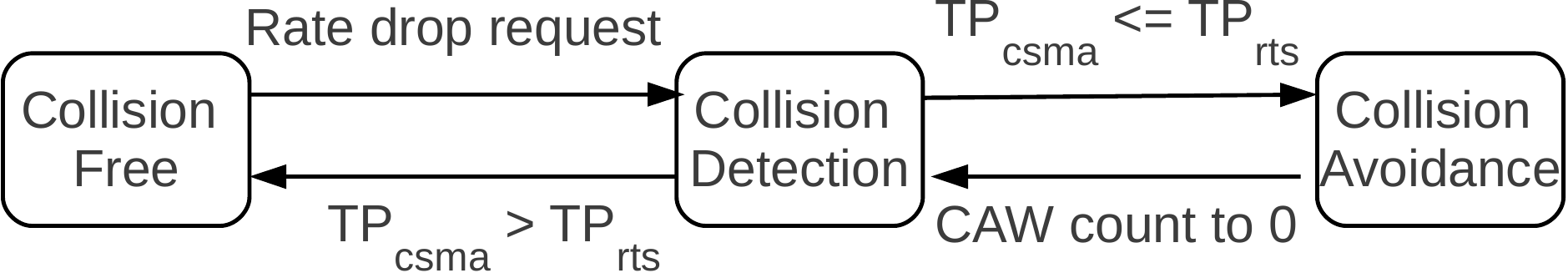}
\caption{State machine of the proposed mechanism.}
\label{fig:statemachine}
\end{figure}

Fig. \ref{fig:statemachine} shows a state diagram for the collision-aware mechanism. When the wireless channel is collision free, data frames are sent according to the retry preference defined in the MRR; that is, within a RAP frames retry at a rate (e.g., $r_{i}$) for $c_{i}$ times before attempting other lower rates. After every RAP the throughput statistics of all supported rates are calculated (for both with or without RTS/CTS) to work out the best performed rate for the next RAP. For each frame, RTS/CTS is turned on or off for each rate in the MRR chain, but in the rate adaptation period, many frames are transmitted with different MRR settings. Retry rates turn on  RTS/CTS with a probability as discussed above. Therefore from a rate perspective,  four pieces of information are gathered including attempts with RTS/CTS on, successes with RTS/CTS on, attempts with RTS/CTS off, successes with RTS/CTS off. Based on this information, for both with or without RTS/CTS, throughput is calculated. When a lower rate achieves higher throughput than the first rate (i.e., $r_{0}$), a rate drop request is issued. The throughput of the current rate for both with and without RTS/CTS is compared, and when the throughput for having RTS/CTS is greater than without it (i.e., $TP_{rts} \geq TP_{csma}$) it concludes that collision is occurring and the RTS/CTS mechanism is turned on in the next RAP rather than dropping the rate. The collision avoidance window (CAW) specifies for how long the RTS/CTS mechanism should be used and it is initialised to one when collision is detected and grows exponentially if collisions still exist when CAW expires. The maximum CAW is 16 considering that its large value could cause poor performance when the collision does not last long. As long as the CAW is bigger than zero, the mechanism is in the collision avoidance state. When CAW reaches zero,  the mechanism enters the collision detection state and the RTS/CTS will be turned off for one RAP. At the end of the RAP, the mechanism compares $TP_{rts}$ and $TP_{csma}$. In the case when $TP_{rts}$ is smaller than $TP_{csma}$ there is no collision and rate adaptation follows the normal retry preference as described earlier, Otherwise the CAW is doubled.

The throughput gain and the overhead of the RTS/CTS exchange are considered when deciding on a rate. The computation of $TP_{rts}$  and $TP_{csma}$ is based on Eqn.\ref{eq: minstrel_Thp} but when computing $T_{tx\_perfect}$, overhead (e.g, DIFS, ACK) is considered according to the IEEE 802.11 standard \cite{IEEE80211_standard_2012}. Specifically  the transmission time of the RTS and CTS frames is considered when RTS/CTS is utilized. 

\subsubsection{RCELC}
RCELC rate control mechanism \cite{weilcn2012, Guerin2012} applies an accurate throughput estimation metric, ELC, as shown in Eqn. \ref{eq: elc}, to rate adaptation and picks up the highest throughput rate in every rate adaptation period.
\begin{eqnarray}
ELC_{r_{i}} = \frac{MSDU}{T_{avg\_transaction}^{r_{i}}}
\label{eq: elc}
\end{eqnarray}
where $MSDU$ is the MAC service data unit (i.e., frame size), and $T_{avg\_transaction}^{r_{i}}$ is the average time spent on successfully delivering a frame using rate $r_{i}$ in the last RAP, including all retransmissions. The averaged transaction time is calculated as
\begin{eqnarray}
T_{avg\_transaction}^{r_{i}} = \frac{T_{total}^{r_{i}}}{Total\_successful\_attempt}_{r_{i}}
\label{eq: avgPktTime}
\end{eqnarray}

where $Total\_successful\_attempt_{r_{i}}$ is the number of successful transmission attempts in one RAP using rate $r_{i}$; while $T_{total}^{r_{i}}$ is the total time for all successful (i.e., $T_{success}^{r_{i}}$) and failed (i.e., $T_{failed}^{r_{i}}$) attempts of rate $r_{i}$ in one RAP, that is the sum of time attempted to deliver a frame using rate $r_{i}$. The frame delivery time is calculated (either $T_{success}^{r_{i}}$ or $T_{failed}^{r_{i}}$) according to the CSMA/CA mechanism in the IEEE 802.11 standard. 

To compute the ELC metric, the statistics of frame delivery (including the attempted rates and their respective count) are gathered on the sender side (based on the corresponding ACK frames arriving within the timeout). As it only requires sender-side information, the ELC metric incurs lower overhead than the approaches that require metric information from the receivers \cite{MutFed}. In addition because it relies on per-frame delivery information, change of traffic load in the network has minimal impact on the accuracy of the ELC metric. 

\subsubsection{MiRA}

MiRA \cite{Pefkianakiston2013} is a rate control mechanism designed for  802.11n networks which support both spatial-diversity-oriented-single-stream (SS) and spatial-multiplexing-driven-multiple-stream (double-stream, DS, in their platform). The SS mode means there is only one stream in transmission and the DS mode means there are two streams being transmitted over the air simultaneously. So the 802.11n rates can be classified into the SS rates and DS rates. A metric called SubFrame Error Rate (SFER) is proposed to study the characteristic of the 802.11n rates, which is defined as 
\begin{eqnarray}
SFER = \frac{nFrames*retries + nBad}{(retries+1)*nFrames}
\label{eq: mira_sper}
\end{eqnarray}
where $nFrame$ is the number of subframes in the transmitted aggregated frame, $nBad$ is the number of subframes received with errors and $retries$ is the number of hardware retries. The observation from extensive experiments is that $SFER$ increases as rates increase when only the SS rates or DS rates are considered. But When looking at the SS rates and DS rates together, the monotonicity does not hold.

MiRA uses a zigzag scheme to opportunistically probe the rate's goodput between the intra- and inter-mode. Within the intra-mode (either the SS or DS rates), it probes the goodput of rates higher and lower than the current rate until it reaches a rate with maximum goodput. When switching to the inter-mode, the probing mechanism only starts if significant changes occur in the measured moving average goodput of the current rate. The goodput is calculated as shown in Eqn. \ref{eq: mira_goodput}
\begin{eqnarray}
G(t) = \frac{DATA*A(t)(1-SFER)}{T_{overhead}+\frac{DATA*A(t)}{R}}
\label{eq: mira_goodput}
\end{eqnarray}
where DATA is the payload size of a MAC-layer frame, $A(t)$ is the moving average of aggregation level (in terms of moving average number of frames). R is the bit rate. $T_{overhead}$ is the various 802.11 communication overhead including DIFS, SIFS and BlockACK.

MiRA also designs a collision aware mechanism to address the hidden terminal problem. It determines a collision loss occurred, if the aggregate frame has experienced at least one retry and the loss ratio of its subframes is less than 10\%.

An adaptive window is used to detect collisions as well. The adaptive window is initialized to 3. For any subsequent frame in the adaptive window, if it satisfies the condition again, MiRA confirms the existence of collisions and triggers the adaptive RTS/CTS mechanism to address collisions. Otherwise, the adaptive window will be decreased by one for each frame that does not satisfy the condition.  RTS/CTS will be turned on only when $\frac{AFRAME}{R} \ge kT_{RCTS}$, where $AFRAME$ is the aggregate frame size, $R$ is the bit rate of rate R, $T_{RCTS}$ is the transmission time of RTS and CTS frames, and $k$ is a benefit cost ratio, set to 1.5 in the prototype. $k$ represents the minimum number of collisions needed to compensate the cost of turning on RTS/CTS. If $\frac{AFRAME}{R} < kT_{RCTS}$, the adaptive window will be reset to zero and RTS/CTS is not triggered.

It starts probing only when the goodput of the current rate changes significantly. This may not be a good reaction. If the channel condition degrades, the goodput of the current rate will be decreased sharply. This probing reaction works fine. However, in a scenario where the current rate is the optimal rate for a long time and the frame loss ratio of the current rate is very low,  if the channel condition improves, the goodput of the current rate will not change significantly. So MiRA will not probe rates and it will lose opportunities to improve throughput. In addition, comparing $\frac{AFRAME}{R}$ and $kT_{RCTS}$ does not directly consider the benefits and drawbacks of the RTS/CTS mechanism so it cannot guarantee that turning off RTS/CTS could provide a throughput gain. The benefit of RTS/CTS is mitigating frame losses and the drawback of RTS/CTS is the RTS and CTS transmission overhead. A better approach could compare these two.

\subsection{Based on signal-to-noise ratio (SNR)}


\subsubsection{Receiver-based AutoRate (RBAR)}
\label{para: RBAR}

In RBAR \cite{Holland2001}, the RTS frame format is modified to carry the modulation rate and the size of the data frame instead of the duration of the channel reservation. Neighboring nodes calculate the duration of channel reservation based on the rate and frame size information. Upon receiving an RTS frame, the receiver selects the appropriate transmission rate for the next data frame to be used by the sender based on the SNR measurement of the RTS frame. A simple threshold based technique is adopted by the receiver to select an appropriate transmission rate, in which each rate has a lower SNR threshold and an upper SNR threshold. The rate selection is based on the comparisons of the estimated SNR and these thresholds. If the measured SNR falls between a rate's upper SNR threshold and lower SNR threshold, the rate is optimally chosen.

The receiver puts the selected rate into a CTS frame along with the data frame size, and transmits the CTS frame back to the sender. Neighboring nodes overhear the CTS frame and calculate the duration of channel reservation. Thus the sender's neighbors may have a different value of the channel reservation duration from the receiver's neighbors. Accounting for this, RBAR includes the final channel reservation in a special sub-header of data frames, called the Reservation SubHeader (RSH).

The modification to RTS/CTS control frames has two goals. One is to provide a mechanism by which the receiver can communicate the selected rate to the sender. The other is to provide neighboring nodes with enough information to calculate the duration of the requested channel reservation. However, the modification of the  RTS and CTS frames causes compatibility problems to the standard and the necessity of the RTS/CTS functionality introduces significant overhead, which is usually disabled in deployment. RBAR is compared with ARF in \cite{Holland2001} and it shows better performance than ARF in a static channel and Rayleigh fading channel conditions.

Another SNR based mechanism that also requires changes in the standard is MutFed \cite{MutFed}. In MutFed, the receiver measures SNR of frames that it receives from a sender. Every tenth frame, the receiver selects a rate by looking up a table based on SNR measurement and notifies it to the sender by transmitting an ACK at the proposed rate. However, the 802.11 standard says that ACKs must be transmitted at mandatory rates. Apart from this, it is hard to synchronize the receiver and the sender, because the number of frame attempts at the sender is not equal to the number of frames received at the receiver if frame losses exist. Therefore a sender may not be able to identify which ACK is the new rate notification except with additional information in the ACK, which again violates the standard. 

RAM \cite{xichen2012} is also a receiver based mechanism. To avoid the synchronization problem, the receiver uses the transmission rate of ACK to notify the sender about the rate for the next frame rather than the tenth frame. The same problem for RAM is that it modifies the ACK transmission rate. Another problem is there may be a long period between the ACK of the last data frame and the next data frame so that the channel quality has already changed during that period of time.

\subsubsection{CHARM}

RBAR and MutFed rely on information collected at the receiver and use existing mechanisms, such as CTS or ACK packets to piggyback the rate information to the sender. Added information requires a change in the standard. CHARM \cite{Judd2007} addresses this problem by allowing the sender to estimate the signal to interference and noise ratio (SINR) metric at the receiver side.

Within a wireless network, neighboring nodes periodically exchange information regarding transmission power and noise level in the form of beacons, probe requests and probe responses. CHARM makes use of this ongoing information to estimate the SINR value at a particular receiver and store the weighted moving average SINR value in a table. Before sending a frame, the sender checks a SINR threshold table for the intended receiver and determines a set of transmission rate. For the first transmission, CHARM selects the highest rate in the set while selecting a lower rate from a fast decreasing rate sequence for retransmissions. Each rate has a minimum required SINR and the threshold is updated by observing frame success rate as a function of predicted SINR. A big advantage for CHARM over RBAR is that it does not require the CTS/RTS mechanism to coordinate rate control and thus eliminates overhead. 

Similar rate control algorithms are SNR Guided Rate Adaptation (SGRA) \cite{SGRA} and LA  (Link Adaptation) \cite{LA}. SGRA uses ACKs' RSSI from a receiver as a predictor of its data frame's RSSI at the receiver side. SGRA uses the measured SNR to map to the frame delivery ratio (FDR) for each rate $R$, then selects the rate which has the maximum throughput computed by $FDR*R$. LA uses the RSSI of beacons and other frames from the receiver. The optimal rate is selected when the measured SNR passes some thresholds. They both assume symmetric link between the sender and receiver, which is not true in wireless channels due to mobility, fading and interference.

\subsubsection{FARA}
Hariharan et al. \cite{Rahul2009} observes that different frequencies have different SNR over the same sender and receiver and there is a potential throughput improvement. FARA leverages the OFDM technology to divide the entire frequency band into subbands. Like RBAR, the SNR metric is monitored at the receiver side. However, a difference from RBAR is that FARA estimates the SNR for each OFDM subband for each sender. In addition, FARA maintains an SNR table which is similar to the one in CHARM. In the table a minimal SNR threshold is defined for each transmission rate. Thus with a measured SNR, FARA can determine a list of supported rates for each subband. The highest transmission rate is selected as the optimal. The link layer ACK frame is modified to carry the selected transmission rate. The novelty of FARA is that it transmits simultaneously multiple frames to a number of next-hops, i.e, one frame per next-hop. This is done by assigning each of these next-hops a non-overlapping OFDM subband. The allocation of frequency is achieved by a randomized greedy approach, which allocates to each next-hop the subband that has better performance for it than other next-hops.

\subsubsection{SNR-aware Intra-frame Rate Adaptation (SIRA)}

%

In  802.11n networks, long aggregate frame is utilized, which enables multiple subframes to be transmitted as a single frame, thus minimizing MAC overhead and improving throughput. However transmitting a longer frame requires longer frame duration and the maximum value would reach 10~ms. This is much longer than the channel coherence time in mobile environments.  Experiments \cite{Hyunjoongicc2014} reveal that subframes located at the latter part of the aggregated frame experiences a higher subframe error rate compared to the former part if the whole aggregated frame is transmitted at the same rate. This motivates SIRA  \cite{Okhwancomletter2015} to transmit the aggregated frame at multiple rates.

The metric used by SIRA is SNR. SNR is estimated based on information extracted from the PHY layer, i.e., the amount of Symbol Dispersion (SD), as the root mean squared value of the difference between the received symbol, $S_r(n)$, and the corresponding transmitted symbol, $S_t(n)$. Specifically, it is calculated according to Eqn. \ref{eq: sira_snr}, where $N$ is the number of symbols in the frame, $E$ is the energy level and $N_o$ is the noise level.
\begin{eqnarray}
SNR =\frac{E}{N_0}\approx \frac{1}{SD^2} = \frac{\frac{1}{N}\sum_{n=1}^N\left|S_t(n)\right|^2}{\frac{1}{N}\sum_{n=1}^N\left|S_r(n)-S_t(n)\right|^2}
\label{eq: sira_snr}
\end{eqnarray}

It is difficult to obtain $SD^2$ from unknown data symbols, so SIRA leverages the known pilot symbol sequence. Each data symbol has a small number of pilot symbols. For example, the number for 20MHz channels is 4. Let $N_p$ represents the number of pilots per symbol. SIRA groups $N_i$ pilots corresponding to $N_s$ data symbols into a received pilot group where $N_i=N_p \times N_s$. Then one aggregated frame is divided into $K$ pilot groups. SIRA estimates SNR for each group $i$, i.e., $SNR_i$ ($1<i<K$) based on Eq. \ref{eq: sira_snr}.  Therefore, SIRA obtains a SNR distribution, SNR= ${SNR_1, SNR_2...SNR_K}$.
 
The notable idea for SIRA is that it selects two transmission rates for a single frame transmission. The primary rate, $R_p$, and the secondary rate, $R_s$. The key point is to find the starting symbol, $I$, where to shift the transmission rate from $R_p$ to $R_s$. In a mobile environment, the estimated $SNR_i$ in SNR monotonically decreases during a frame reception. Using this monotonicity feature of SNR, SIRA detects mobility. SIRA finds $I$ from the beginning of the frame when the condition $SNR_i < SNR_{th}(R_p)$ is satisfied, where $SNR_{th}(R_p)$ is the minimum SNR at which the theoretical BER of $R_p$ is less than $10^{-4}$. Then $I$ is fed back to the sender.

The primary $R_p$ is determined by inter-frame rate control mechanisms, like SGRA. The secondary rate $R_s$ is chosen by table look-up. Each row in the table specifies $R_p$ and the corresponding $R_s$ and $SNR_{th}$. In \cite{Okhwancomletter2015} the performance of SGRA \cite{SGRA} and SGRA+SIRA is evaluated and the results demonstrate that SIRA can improve SGRA's throughput significantly in mobile environments. One limitation of SIRA is that it can only determine two rates for an aggregated frame, which may not be enough for a fast changing channel.

\subsubsection{OAR}
The OAR mechanism \cite{Sadeghi2002} is an extension for the MAC layer rate control mechanisms, in which it sends multiple back-to-back data frames whenever the channel quality is good. It observes that channel coherence time (duration for which the channel condition of a mobile station is better-than-average) is typically larger than multiple frame transmission time. Thus it enables the sender to hold the channel for transmissions for an extended time. As illustrated from experiments in \cite{Sadeghi2002}, RBAR with the OAR extension achieves significant throughput gains compared with RBAR. However, this may introduce unfairness problems as a station can transmit multiple frames once it detects a good channel.


\subsection{Based on effective SNR}
\subsubsection{ESNR}
SNR does not capture the effect of frequency-selective fading. ESNR \cite{Halperinsigcomm2010} computes the effective SNR to address this issue. In 802.11n networks, OFDM (Orthogonal Frequency Division Multiplexing) technology is used. In these networks, 20 MHz or 40 MHz channels are divided into 312.5 kHz bands named subcarriers. Each of the subcarriers sends independent data simultaneously. Each subcarrier in a frame is modulated equally, using BPSK, QPSK, QAM-16 or QAM-64, with 1, 2, 4 or 6 bits per symbol, respectively. The transmission rate depends on the modulation and coding.  

Effective SNR is not the average subcarrier SNR. The 802.11n Channel State Information (CSI) is used to calculate the effective SNR. The CSI is a collection of M*N matrixes $H_s$. Each matrix represents the RF path (SNR and phase) between all pairs of $N$ transmit and $M$ receive antennas for one subcarrier $s$. Let $SNR_s$ be the SNR for each subcarrier $s$. ESNR calculates the average BER across all subcarriers as displayed in Eqn. \ref{eq: BER_Carrier}
\begin{eqnarray}
BER_{eff,k} = \frac{1}{52} \sum BER_k(SNR_s)
\label{eq: BER_Carrier}
\end{eqnarray}
where $k$ is the modulation rate. Then ESNR \cite{Halperinsigcomm2010} uses the inverse mapping from BER to get the effective SNR as denoted in Eqn. \ref{eq: effective_SNR}
\begin{eqnarray}
SNR_{eff,k} = BER_k^{-1} (BER_{eff,k})
\label{eq: effective_SNR}
\end{eqnarray}

Either the sender or receiver can perform effective SNR calculation. For a sender to make rate decision, the receiver's effective SNR thresholds for different rates will be sent to the sender during association. The sender uses two methods to get the up-to-date CSI (channel state information). The receiver sends it to the sender  or the sender estimates it from the reverse path. Given the recent CSI, the highest rate, which is predicted to have a frame delivery ratio above 90\%, is selected.

\subsubsection{InFRA}
The novelty of  InFRA \cite{Hyunjoongicc2014} is that a frame is divided into multiple  groups of symbols (GOS) and the sender selects the transmission rate for each GOS. 
\begin{table}[tbh]
\caption{Bit Error rate vs. SNR $\gamma$}
\centering
\label{tab:BER_SNR}
\begin{tabular}{ c p{2.1cm} p{1.8cm} p{1.8cm}}
\toprule
Modulation & Bits/Symbol & $BER_k(\gamma)$ \\
\midrule
BPSK & 1 & $Q(\sqrt{2\gamma})$ \\
QPSK & 2 & $Q(\sqrt{\gamma})$ \\
QAM-16 & 4 & $\frac{3}{4}Q(\sqrt{\frac{\gamma}{5}})$ \\
QAM-64 & 6 & $\frac{7}{12}Q(\sqrt{\frac{\gamma}{21}})$ \\
\bottomrule
\end{tabular}
\end{table}
In InFRA, each GOS includes 5 OFDM symbols. Suppose the estimated SNR at the $n^{th}$ subcarrier of $s^{th}$ symbol in a GOS is $\gamma_{s,n}$. Let $BER_{eff,k}$ denote the average BER for the GOS using transmission rate $r_k$. The transmission rate in 802.11n depends on the modulation, coding and the number of streams. $BER_{eff,k}$ is calculated as
\begin{eqnarray}
BER_{eff,k} =\frac{1}{5N_{sc}}\sum_{s=1}^5\sum_{n=1}^{N_{sc}}BER_k(\gamma_{s,n})
\label{eq: infra_brr}
\end{eqnarray}
where $N_{sc}$ is the number of subcarriers in an OFDM symbol. $BER_k(\gamma)$ is a bit error rate function using transmission rate $r_k$ for SNR $\gamma$. Table \ref{tab:BER_SNR} displays $BER_k(\gamma)$ as a function of $r_k$, where $Q$ is the standard normal CDF \cite{goldsmith2005wireless}. Then using a inverse mapping function, the measured BER is mapped into the effective SNR, $\gamma_{eff}$.
\begin{eqnarray}
\gamma_{eff} =BER^{-1}(BER_{eff,k})
\label{eq: infra_effsnr}
\end{eqnarray}
Then InFRA selects $r_i$ as the best rate for next GOS when $\beta_i < \gamma_{eff} < \beta_{i+1}$ where $\beta_i$ is the minimal SNR for transmission rate $i$. 

These GOSs are transmitted to the receiver sequentially. The receiver decodes every GOS and uses CRC checksum embedded in the pilot subcarriers of the GoS to determine whether it is delivered successfully. An ACK/NACK will be sent via a separate  channel to the sender depending on whether a GOS is successful or not. Negatively acknowledged GOS will be retransmitted until it is successful. 

After decoding each GOS, the receiver selects the best transmission rate based on the calculated SNR and sends it back to the sender via ACK or NACK. After all GOSs are received successfully, the receiver sends the end-of-stream (EOS). This method avoids frame-level retransmission, and therefore reduces retransmission overhead. However, in order for the receiver to decode the GOS correctly, InFRA encodes six bits of the transmission rate into the GOS header. This is a type of added communication overhead. Also, the requirement of using another channel for ACK/NACK also consumes precious frequency resources.

\subsubsection{TurboRate}
TurboRate \cite{weiliang2014} is a rate control mechanism proposed for uplink (client to Access Point) transmissions in  multi-user MIMO networks. By listening to Access point (AP)'s transmissions, the client calculates two variables: (1) the direction along which the client's signals arrives at the AP, and (2) its SNR at the AP if it is transmitted alone.  When multiple clients contend for transmission, the client which wins the contention first includes into the frame the information about the direction along which the AP receives its signal. A client that wants to transmit simultaneously with the first client uses the information to project its signal orthogonal to the first client and calculates the reduction in its SNR. After estimating the SNR after projection, the client computes the effective SNR \cite{Halperinsigcomm2010}. It then maps the effective SNR to the transmission rate using SNR-bitrate tables \cite{Rahul2009}. Additionally, the client concurrently calculates the effective SNR and selects the rate using the same method.

\subsection{Based on bit error rate (BER)}
\subsubsection{SoftRate}
SoftRate \cite{Vutukuru2009} estimates the bit error rate (BER) metric, which is calculated over each received frame, to adjust the transmission rate.

The average BER of the channel at the current transmission rate $R_{current}$ is calculated over all bits in a frame $BER(R_{current})$, while the prediction of BER at other transmission rates is based on two heuristics. One is that the BER is a monotonically increasing function of the rate at any given SNR. The other is that a rate's BER at a given SNR is at least a factor of 10 higher than that of the next-lower rate \cite{Vutukuru2009}.

Based on the two heuristics, SoftRate calculates BER for multiple rates and computes optimal thresholds for each rate. For each rate, SoftRate calculates the lower and upper limit, $BER_{lower}$ and $BER_{upper}$ respectively. The SoftRate selects the next rate $R_{next}$ as
\[
    R_{next}= 
\begin{cases}
    R_{current}^{-}, & \text{if } BER(R_{current}) \ge BER_{upper}\\
    R_{current}^{+}, & \text{if } BER(R_{current}) \le BER_{lower} \\
    R_{current},        & \text{otherwise}
\end{cases}
\]


In addition, if the measured BER is far from the lower or upper limits, SoftRate jumps multiple levels to improve the responsiveness of significant change in channel conditions. However, the details of rate jump are not described in \cite{Vutukuru2009}. The BER metric is measured at the receiver and sent back to the sender to assist with the rate decision. This requires modification to the MAC-layer frames and can cause compatibility issues.


\subsubsection{AARC}
A similar metric to BER, A posteriori Bit Error Probability (ABEP), is proposed in \cite{lixingICC2015} and a rate control mechanism called AARC that is based on this metric is designed. ABEP  is the error probability of each bit based on the received wireless signal symbols. 

For a received frame, the receiver calculates the ABEP metric. If the frame is received correctly, the receiver decides whether to increase/decrease or maintain the rate based on comparison of the measured ABEP and two thresholds. These two thresholds are defined for each rate separately. If the measured ABEP is above the higher threshold, the rate is decreased. If the calculated ABEP is below the lower threshold, the rate is increased. Otherwise, the rate is unchanged. The rate decision is notified back to the sender. If the frame is received incorrectly, the receiver determines the reason whether it is due to collisions or fading and the reason is notified to the sender. The way to differentiate collisions and fading is ABEP characteristics. For collisions, ABEP will experience an upsurge when collisions start and the upsurge will last for a relatively long period until the collisions finish. Slow fading has no upsurge. Fast fading has an upsurge but it does not last long.

\subsection{Based on FER}

A rate control mechanism \cite{ShamyGlobecom2008} is proposed to support the IEEE 802.11e quality of service (QoS). Rate adaptation is conducted according to the FER (Frame Error Rate) metric \cite{LeeICNS2006}. The FER metric is calculated based on measured SNR and BER, as shown in Eqn. \ref{eq: BER}.
\begin{eqnarray}
FER = 1-(1-BER*SNR)^N
\label{eq: BER}
\end{eqnarray}
where N is the number of bits of the received frame. It increases the transmission rate when the number of consecutive ACKs reaches ten. It decreases the current rate $r_h$ to its next lower rate $r_l$, when the measured FER exceeds the threshold $e_h$. $e_h$ is computed as $e_h = 1- \frac{r_l}{r_h}(1-e_l)$ where $e_l$ is the FER of $r_l$. This guarantees that decreasing the rate will have a higher throughput than staying at the higher rate. In order to satisfy the delay requirement, the rate control mechanism calculates the maximum retransmission limit $L$ based on the delay requirement $D$, the contention window  size, $CW$, and the average slot time $t$. The maximum retransmission limit is computed according to $L=\log_2(\frac{2D}{CW*t}+1)-1$.

This mechanism gradually increases and decreases the transmission rate so it may respond slowly to channel improvement and degradation. The channel quality metric calculation is more complex than any mechanism that is based on a sole metric. To obtain the FER metric, SNR and BER need to be calculated first. But this may imply that the mechanism is more accurate in channel estimation.

\subsection{Based on combined metrics}
In contrast to the mechanisms we have described so far, HARC \cite{HARC} bases its rate control decisions on multiple metrics, including throughput, FLR and SNR. The key component of HARC is a throughput based controller. The controller generates the optimal rate and decides whether to probe adjacent rates. The rate decision made by the core controller can be overwritten by a SNR based controller. For each rate, the SNR controller measures the Signal Strength indication of the Acknowledged frames (SSIA). The SNR based controller has a look-up table. In this table, each rate has three SSIA thresholds, of which two are for stable channel conditions (a high and a low threshold) and an additional low threshold is for dynamic link conditions. The low threshold for the stable channel refers to the one which can provide acceptable performance (defined as FLR $<$ 10\%). The dynamic channel condition is detected by comparing the last three consecutive SSIA values, $SSIA_1$, $SSIA_2$ and $SSIA_3$. If both $SSIA_2-SSIA_1$ and $SSIA_3-SSIA_2$ exceed a predefined value, a dynamic channel is detected and the corresponding low threshold is used. Otherwise the stable channel low threshold is used. Both the static and dynamic channel use the same high threshold.
In HARC, there are other two threshold rates, $R_{upbound}$ and $R_{lowbound}$, which provide a bound for the minimum and maximum rate to be used. Their values are determined by the corresponding SSIA values. If the newly selected rate is below the $R_{lowbound}$, $R_{lowbound}$ is used as the new rate, vice versa for the $R_{upbound}$. 

In \cite{HARC} authors do not compare HARC with other rate control mechanisms. However, HARC is based on many thresholds and parameters. We conjecture the ability to find a set of parameters that will allow HARC to achieve maximum performance in all scenarios. We learned from the PID algorithm that this task is challenging.

%% file: analysis.tex
\section{Analysis and Discussion}
\label{analysis}

In this section, we  compare the rate control algorithms in five aspects: (1) metric selected by the mechanism, (2) how the rate is updated - gradual adaptation or best rate selection, (3) where the quality evaluation and rate decision is made, at the sender or the receiver side, (4) when the rate is updated -  per-frame adaptation or based on rate adaptation period, and (5) what the rate adaptation algorithm is. The comparisons are summarized in Table \ref{tab:Eval_MAC_layer}.


%

\begin{landscape}
\begin{longtable}[l]{ l c c c c p{8.8cm} }

\caption{Classification of MAC-layer rate control mechanisms} \\

\toprule 
	Mechanism    & network  &  \multicolumn{3}{c}{Classification factors}  &  Adaptation strategy   \\ 
				 & Metric (Where) &  How  & Duration &      \\
\midrule
AARC \cite{lixingICC2015} & 802.11abg & ABEP (S) & Gradual & Per-frame & $R^{\uparrow}$ if $ABEP(R_{current})< threshold_{lower} $, $R^{\downarrow}$ if $ABEP(R_{current}) > threshold_{upper}$ \\
AARF \cite{Lacage2004a}& 802.11abg & CTR (S) & Gradual & Per-frame & Same as ARF, dynamically adapt the $\alpha$ threshold to achieve convergence \\
AMRR \cite{Lacage2004a} &802.11abg & FLR (S) & Gradual & RAP &  $R^{\uparrow}$ if $FLR < 10\%$, $R^{\downarrow}$ if $ FLR > 33\%$  \\
ARF \cite{Karmerman1997} &802.11abg & CTR (S)  & Gradual & Per-frame  & $R^{\uparrow}$ if $C_{success}\ge 10$, $R^{\downarrow}$ if $C_{fail}\ge 2$ \\
BEWARE \cite {BEWARE} &802.11abg & $T_{tx}$ (S) & Best rate & RAP  & Select rate $R$, such that $min(T_{tx}(R))$ \\
CARA \cite{CARA} &802.11abg & CTR (S) & Gradual & Per-frame & Improve ARF with RTS probing and CCA detection to address collisions  \\
CHARM \cite{Judd2007} & 802.11abg& SNR (S) & Best rate & Per-frame & Using measured SNR to find all supported rates from a table listing the minimal SNR for each rate and select the highest one\\
ERA \cite{shaoentechrep2007}&802.11abg & CTR (S) & Gradual & Per-frame & Same as AARF but uses fragmentation to address hidden terminal collisions \\
ESNR \cite{Halperinsigcomm2010}& 802.11n & Effective SNR(R) & Best rate & Per-frame & Given  measured effective SNR, computes the highest rate of which the packet delivery ratio is greater than 90\% \\
FARA \cite{Rahul2009}& 802.11abg & SNR (R) & Best rate & Per-frame & Using measured SNR to find all supported rates from a table listing the minimal SNR for each rate and select the highest one  \\
HA-RRAA \cite{Ioannis2013}& 802.11abg & FLR (S) & Gradual & RAP & Same as RRAA, but prolongs the time a rate is used when selected to avoid rate oscillation, uses a shorter window to monitor FLR to make it responsive to channel degradation, and considers RTS overhead to enhance efficiency in addressing collisions. \\
HARC \cite{HARC}&802.11abg & SNR, TP (S) & Best rate & RAP & Select rate $R$, such that $max(TP(R))$ while using measured SNR to verify whether it is applicable \\
InFRA \cite{Hyunjoongicc2014}& 802.11n & Effective SNR (R) & Best rate & Per-frame & Different rates will be used for different parts, called GOS, of a frame. Using measured effective SNR,$\gamma_{eff}$, it selects $r_i$ as the best rate for next GOS when $\beta_i < \gamma_{eff} < \beta_{i+1}$ where $\beta_i$ is the minimal SNR for transmission rate $i$  \\
LA \cite{LA} & 802.11abg& SNR (S) & Best rate & Per-frame & The optimal rate is selected when the measured SNR passes some thresholds \\
Minstrel \cite{minstrel} & 802.11abg& TP (S) & Best rate & RAP & Select rate $R$, such that $max(TP(R))$  \\
Minstrel-rts \cite{mac80211ratecontrol2015} & 802.11abg& TP (S) & Best rate & RAP & Same as Minstrel, but added RTS probing to address hidden terminal problems \\
MiRA \cite{Pefkianakiston2013} & 802.11abg& TP (S) & Best rate & RAP & Select rate $R$, such that $max(TP(R))$  \\
MutFed \cite{MutFed} & 802.11abg& SNR (R) & Best rate & Per-frame & Rate selection is based on a table lookup, where the rate selected is most appropriate at the given average SNR \\
Onoe \cite{Onoe} &802.11abg & FLR (S) & Gradual & RAP & $R^{\uparrow}$ if credit $\ge 10$, $R^{\downarrow}$ if FLR $>$ 50\%  \\
PID \cite{PID} & 802.11abg& FLR (S) & Best rate & RAP & $R_{current} \leq R_{new} \leq (R_{current}+adj)$ if $adj>0$; $(R_{current}+adj)\leq R_{new}\leq R_{current}$ if $adj<0$ \\
PIDE \cite{MSWiM2011}&802.11abg & FLR, TP (S) & Best rate & RAP & Same as PID, but switching rate only if $TP(R_{new}) > TP(R_{current})$ \\
RAM \cite{xichen2012}&802.11abg & SNR (R) & Best rate & Per-frame & Maintain a throughput-vs-(rate, SNR) table. Using measured SNR, RAM looks up the table and selects the rate R that can maximise throughput.\\
RBAR \cite{Holland2001}&802.11abg & SNR (R) & Best rate & Per-frame  & Select rate $R$, such that  $threshold_{upper}>SNR(R)>threshold_{lower}$  \\
RCELC \cite{weilcn2012}&802.11abg & TP (S) & Best rate & RAP & Select rate $R$, such that $max(TP(R))$  \\
RRAA \cite{Wong2006}& 802.11abg& FLR (S) & Gradual & RAP & $R^{\uparrow}$ if $FLR < threshold_{lower}$, $R^{\downarrow}$ if $FLR > threshold_{upper}$  \\
SampleRate \cite{Bicket2005}& 802.11abg& $T_{tx}$ (S) & Best rate & RAP & Select rate $R$, such that $min(T_{tx}(R))$  \\
SGRA \cite{SGRA}& 802.11abg& SNR (S) & Best rate & Per-frame  & Use measured SNR to get the frame delivery ratio (FDR) for each rate $R$, then choose the rate which has the maximum throughput computed by $FDR*R$ \\
SIRA \cite{Okhwancomletter2015}& 802.11n & SNR(R) & Best rate & Per-frame & Two rates will be chosen for two parts of a frame from looking up a table with each row listing two rates and correspondent SNR \\
SoftRate \cite{Vutukuru2009}&802.11abg & BER (S, R)  & Best rate & Per-frame & $R^{\uparrow}$ if $BER(R_{current})< threshold_{lower} $, $R^{\downarrow}$ if $BER(R_{current}) > threshold_{upper}$ \\
TurboRate \cite{weiliang2014}& 802.11ac & Effective SNR (S) & Best rate & Per-frame & Using measured effective SNR to find all supported rates from a table listing the minimal effective SNR for each rate and select the highest one\\
\bottomrule

\label{tab:Eval_MAC_layer}

\end{longtable}
\end{landscape}

\subsection{Metric}

Various metrics are utilised in rate control mechanisms including CTR, FLR, transmission time, throughput, SNR, effective SNR, BER and combined metrics. For the CTR and FLR metrics, the calculation is simple, and only some registers are required to maintain the transmission status. They are updated based on the transmission failure/success of the last frame. One drawback of these metrics is that the link bandwidth information is not taken into account. This indicates that high rates will not be preferred in collision environments where all rates have high losses, which limits the performance. Transmission time and throughput are a bit more complex than CTR and FLR, knowledge of the physical layer features is required including symbol interval, PLCP header preamble duration and bit rate of the rate.  The bandwidth and loss ratio of the link are considered, which are two important aspects of the link condition. Therefore, even with a high loss ratio, the high rate will be selected if it can achieve the best throughput or lowest transmission time. SNR is not directly connected to performance. Some research findings show that SNR is directly related to loss ratio in controlled environments but this does not hold true in realistic environments. This means SNR is not a good indicator for throughput. Effective SNR is a new metric and the calculation requires BER information which is not connected to throughput as well. For example, in a collision environment, a collided frame can have a very high BER but a single frame loss has little impact on the aggregate throughput.

According to the strategy how link quality metrics are gathered, rate control mechanisms can be categorised into 3 classes. The first is a probing and testing based approach which sends a number of data frames at an attempted rate other than the current rate to test how the attempted rate works, such as ARF, AARF. The benefit of this approach is that these rate control mechanisms know how two of these available rates (current and adjacent higher rate) perform. However it lacks a global view of the rate performance. In some cases, e.g.,  802.11n networks \cite{Pefkianakiston2013}, the adjacent rate may not perform better than the current rate but other rates may do. This approach could fail in these cases. The second scheme is a non-probing approach which  uses transmission results of the current rate to predict an appropriate transmission rate for the future, such as PID, RRAA and SoftRate. The problem with this approach is that it should have a mechanism to predict how other rates perform. Otherwise blindly increasing the rate could cause the rate oscillation problem \cite{MSWiM2011}. The third approach is a hybrid approach, which is not only based on the transmission results of the current rate, but also uses probing data frames to test the network conditions, such as Minstrel. The benefit of this approach is that it has a global view of how all rates perform so that it is robust in collision environments \cite{mac80211ratecontrol2015}. The problem for this approach is that it uses a number of trial frames to be transmitted at some rates other than the best throughput rate, which adds overhead.

\subsection{Adaptation Decision}

According to where the link quality metric is measured and the rate decision is made, rate control mechanisms are classified into three groups: sender, receiver or hybrid based approaches. 

CTR, FLR, transmission time and throughput based mechanisms including ARF, RRAA and Minstrel are in the sender based group, because these metrics are calculated based on received acknowledgements. Most of the SNR based mechanisms, e.g., RBAR, are receiver based to enhance the measurement accuracy. However to get the rate feedback back to the sender requires modification of the current frame format, which violates the standard. Some SNR based mechanisms are sender based, e.g., CHARM. In such mechanisms, the sender predicts SNR based on measured RSSI and transmission power indicated in the probing packets. Those mechanisms assume an asymmetric link between communication pairs. Unfortunately, it is not true in many realistic environments. Another group is a hybrid approach in which the metric estimation and rate selection are made on different sides. SoftRate is such an example. The receiver estimates BER which is returned to the sender for rate selection.

\subsection{Algorithm}

According to how the rate is adapted, we can classify rate control mechanisms into two schemes. The first strategy is called a gradual rate adaptation (or \emph{rate ladder}) scheme in which the transmission rate is increased or decreased to the next level, whereas the second is the best rate selection scheme which  directly selects the best rate. 

As shown in Table IV, most of the CTR and FLR based mechanisms including ARF, AARF, Onoe, AMRR, and RRAA take the first scheme and attempt to upgrade to the next higher lever rate whenever the channel is good. For CTR based mechanisms, a good channel is inferred by 10 consecutive successful transmissions while for FLR based mechanisms it means the measured FLR is less than a threshold. If the channel condition is bad, gradual rate adaptation mechanisms decrease to the next lower lever rate. a bad channel is detected by 2 consecutive frame failure or when the measured FLR is bigger than another threshold. SNR, BER, transmission time and link capacity estimation based mechanisms including SampleRate, RBAR and SoftRate adopt the second method and are able to select the best rate in only one step.

The best rate selection is more responsive than gradual rate selection when channel suddenly changes or the rate control mechanism starts or restarts. However, in order to achieve best rate selection, a percentage of frames (10\% in SampleRate and Minstrel) need to be attempted at other rates to gather fresh measurement. This gains rate control mechanisms a full view about how well all transmission rates can perform. 

Some of the best rate selections are likely to perform better than gradual rate selections in a hidden terminal environment, because high frame losses caused by collisions trigger those mechanisms to decrease the rate to the lowest rate. After that, gradual rate selection mechanisms (FLR based) stick to the lowest rate and cannot increase rate again due to high FLR. Usually a higher rate performs better than a lower rate in hidden terminal environments because the higher rate has less channel air time and less collision probability. For the link capacity based best rate selection mechanisms, e.g., Minstrel, it is likely that  the best throughput rate is selected \cite{mac80211ratecontrol2015}, as fresh statistics of all rates are available.


%% file: environments.tex
\section{Experimental evaluation}
\label{environments}


In this section we briefly describe  evaluation environments (simulation, over-the-air, and controllable platforms)  used for evaluation of MAC layer rate control algorithms and also show which rate control approaches were evaluated/compared in particular evaluation environments.
 
In simulation, a network simulator, e.g. NS3 \cite{NattavitComComAp2012} is usually utilised to simulate and predict the behavior of a network without the presence of an actual network. It is easy to write scripts based on the simulator's APIs to simulate network scenarios. In addition, network simulations are also inexpensive, as some simulators, eg. NS2/NS3, are free to use and still some simulators, e.g. Qualnet \cite{Comparetto2011} provide a trial version for academics. Using simulations saves funds which can be used to buy hardware like computer and cables. Additionally, simulators provide engineers and researchers with the means to test network scalability that might be extremely difficult using real hardware, for instance, simulating a large network with hundreds of computer nodes. Another advantage of simulations is that the results are repeatable. Using the same configuration and parameter setting, experimental results can be reproduced. However, simulations also have some drawbacks. For example, there are no actual stations or wireless signals in the simulated networks and the simulation is based on very simple noise models. Those models cannot accurately simulate the real environment dynamics, such as movement of obstacles (e.g. moving vehicles) and presence of various types of interference (e.g. microwave).

Many rate control mechanisms are available on the NS3 platform, including AARF, AMRR, ARF, CARA, RRAA, Minstrel, etc. NS3 cannot simulate some hardware features. For example, NS3 does not provide a data structure for the multi-rate-retry chain. In the mac802.11 framework, a Linux Kernel component, which integrates common features, e.g., rate control, to simplify a wireless driver development, such a data structure is provided to notify the driver which rates will be utilised if retransmission is needed. After finishing transmission, the driver updates the data structure and notifies the mac802.11 framework which rates are actually attempted and how many times, respectively. Based on this information, the mac802.11 framework can calculate statistics for the metric of a rate control mechanism. It is not difficult to set the rate-retry-chain on the NS3 platform, but it is a tough task to obtain the transmission status after transmission because there is no driver at the lower layer to update the data structure. The current approach for the Minstrel implementation regards the retransmission times of all rates in the multi-rate-retry chain as the retransmission count for the first rate in the chain. Therefore, the retransmission status for lower rates in the chain is neglected. This, without doubt, negatively affects  the performance evaluation of the Minstrel mechanism. 

Compared to simulations, the deployment of an over-the-air experiment platform (e.g. ORBIT \cite{Raychaudhuri2005}), is more expensive particularly when building a large network. It is also difficult to maintain it. Such maintenance work includes trouble-shooting with link breaks and security problems. In over-the-air experiments, the wireless signal travels over the air so it captures the real environment dynamics caused by movement of obstacles and various types of interference from competing stations, microwave ovens and so on. However, it is hard to reproduce the over-the-air results because the environment conditions vary. 

As discussed, wireless network simulations are based on very simple noise models and rate control mechanisms also could not be fairly compared in over-the-air experiments. Controllable platforms \cite{BonneyAero2008}\cite{Bialkowski2010} (also called emulation) are designed to solve these problems. On these platforms, the wireless signal travels over co-axial cables instead of over-the-air and a variable attenuator is utilised between a traffic source and sink to generate various link conditions. All stations are put into separate shielding boxes to isolate stations from external interference, therefore performance is compared fairly. However, the drawback of a controllable platform is that it can only construct a small network for evaluation so the contention environment may not be created using controllable platforms.

The performance comparison based on simulations, emulation platforms and over-the-air experiments are summarised in Table \ref{tab:exp_MAC_layer}.

\renewcommand{\multirowsetup}{\centering} 
\begin{landscape}
\begin{longtable}[l]{c c c c}
\caption{Performance Comparison As Reported in Literature} \\
\hline 
\toprule
Env & Experiment Settings  &  Comparison ($A<B$: B better than A)  & Ref   \\
\toprule

\multirow{3}{1cm}{SIM}
     & slow fading & SampleRate $<$ RRAA $<$ CHARM $<$ RBAR $<$ \underline{SoftRate} &
     \multirow{3} {1cm} {\cite{Vutukuru2009}} \\ 
    & fast fading & RBAR $<$ SampleRate $<$ RRAA $<$ \underline{SoftRate}  &\\
    & hidden terminal & RRAA $<$ \underline{SoftRate}(\emph{nid}) $<$ SampleRate $<$ \underline{SoftRate}(\emph{id})  & \\
\hline

{SIM}
    & slow fading &  Minstrel $<$ RRAA $<$ CARA $<$ \underline{AARC} & \cite{lixingICC2015} \\
\hline

{SIM}
    & mobile &  ARF $<$ AARF $<$ \underline{MutFed} & \cite{MutFed} \\
\hline

\multirow{2}{1cm}{SIM}
     & vary contending stations &  ARF $<$ ARF-RTS $<$ \underline{BEWARE} $<$ CARA &
    \multirow{2} {1cm} {\cite{BEWARE}} \\
     & ricean fading &  ARF $<$ CARA $<$ ARF-RTS $<$ \underline{BEWARE} &\\
\hline

{SIM}
    & walking speed from 0 to 1m/s &  SGRA $<$ SGRA+\underline{SIRA} & \cite{Okhwancomletter2015} \\
\hline

{SIM}
    & static, mobile, contention &  ARF $<$ CARA $<$ ERA & \cite{saadiscc2008} \\
\hline

\toprule

\multirow{4}{1cm}{EMU}
     & walking speed at 0.5m/s &  Onoe $<$ SampleRate $<$ AMRR  $<$ \underline{CHARM} &
     \multirow{4} {1cm} {\cite{Judd2008}} \\
    & walking speed at 1, 2m/s &  Onoe $<$ AMRR $<$  SampleRate $<$ \underline{CHARM} &\\
    & hidden terminal & AMRR  $<$ Onoe $<$  SampleRate $<$ \underline{CHARM} & \\
\hline

\multirow{2}{1cm}{EMU}
     & human speed mobile &  SoftRate $<$ SampleRate $<$ \underline{ESNR} &
     \multirow{2} {1cm} {\cite{Halperinsigcomm2010}} \\
    & fast mobile & SoftRate $<$ SampleRate $<$ \underline{ESNR} &\\
\hline

{EMU}
    & vary interference load &  PID $<$ PIDE $<$ Minstrel $<$ \underline{Minstrel-rts} & \cite{mac80211ratecontrol2015} \\
\hline

\multirow{4}{1cm}{EMU}
     & static & PID $<$  Minstrel $<$ \underline{PIDE} &
     \multirow{4} {1cm} {\cite{MSWiM2011}} \\
    & channel quality linearly increases &  PID $<$  Minstrel $<$ \underline{PIDE} &\\
    & channel quality linearly decreases &  PID $<$  Minstrel $<$ \underline{PIDE} &\\
     & channel quality sudden changes &  PID $<$  Minstrel $<$ \underline{PIDE} &\\
\hline

\multirow{4}{1cm}{EMU}
     & static &  Minstrel $<$ \underline{RCELC} &
     \multirow{4} {1cm} {\cite{weilcn2012}} \\
    & channel quality linearly increases &   Minstrel $<$ \underline{RCELC} &\\
    & channel quality linearly decreases &   Minstrel $<$ \underline{RCELC} &\\
     & channel quality sudden changes &   Minstrel $<$ \underline{RCELC} &\\
\hline

\multirow{2}{1cm}{EMU}
     & 0.1ms $<$ coherence time $<$ 1ms &  RBAR $<$ RBAR with OAR $<$ ARF $<$ RRAA &
    \multirow{2} {1cm} {\cite{Camp2008a}} \\
     & 1ms $<$ coherence time $<$ 0.1s & RBAR $<$ ARF $<$ RRAA $<$ RBAR with OAR &\\
\hline

\multirow{6}{1cm}{EMU}
     & static &  Onoe $<$ AMRR $<$  Minstrel $<$ SampleRate &
     \multirow{6} {1cm} {\cite{YIN2010}} \\
    & vary interference strength &  Onoe $<$ AMRR $<$  SampleRate $<$ Minstrel &\\
    & vary interference interval &  Onoe $<$ AMRR $<$  SampleRate $<$ Minstrel &\\
    & vary interference duration &  Onoe $<$ AMRR $<$  SampleRate $<$ Minstrel &\\
    & vary hidden terminal load &  Onoe $<$ SampleRate $<$  AMRR $<$ Minstrel &\\
    & vary hidden terminal link quality & Onoe  $<$ SampleRate $<$  AMRR $<$ Minstrel &\\    
\hline

\toprule

\multirow{2}{1cm}{OTA}
     & 100~MHz channel & SampleRate $<$ \underline{FARA} &
    \multirow{2} {1cm} {\cite{Rahul2009}} \\
     & 20~MHz channel & SampleRate $<$ \underline{FARA} &\\
\hline

\multirow{7}{1cm}{OTA}
     & 802.11a, UDP &  RRAA $<$ SampleRate $<$  ARF $<$ \underline{HA-RRAA} &
     \multirow{6} {1cm} {\cite{Ioannis2013}} \\
    & 802.11a, TCP (4 flows) &  SampleRate $<$ ARF $<$  RRAA $<$ \underline{HA-RRAA} &\\
    & 802.11a, TCP (1 flow) &  SampleRate $<$ RRAA $<$  ARF $<$ \underline{HA-RRAA} &\\
    & 802.11a, mobile, UDP &  SampleRate $<$ RRAA $<$  \underline{HA-RRAA} $<$ ARF &\\
    & hidden terminal &  ARF $<$ SampleRate $<$ RRAA $<$  \underline{HA-RRAA}    &\\
    & 2.4GHz channel &  ARF $<$ SampleRate $<$ RRAA $<$  \underline{HA-RRAA} &\\
    & 5GHz channel &  SampleRate $<$ ARF $<$  RRAA $<$ \underline{HA-RRAA} &\\
\hline

\multirow{2}{1cm}{OTA}
    & indoor &  ARF $<$ AARF $<$ Onoe $<$ \underline{SampleRate} &
    \multirow{2} {1cm} {\cite{Bicket2005}} \\
    & outdoor & ARF $<$ AARF $<$ Onoe $<$ \underline{SampleRate} & \\
\hline

\multirow{5}{1cm}{OTA}
     & TCP, static, 802.11a &  ARF $<$ AARF $<$ SampleRate $<$ \underline{RRAA} &
     \multirow{5} {1cm} {\cite{Wong2006}} \\
    & UDP, static, 802.11a & ARF $<$ AARF $<$ SampleRate $<$ \underline{RRAA} & \\
    & UDP, static, 802.11b & ARF $<$ AARF $<$ SampleRate $<$ \underline{RRAA(A-RTS)} $<$ \underline{RRAA} & \\
    & UDP, mobility, 802.11b & SampleRate $<$ AARF $<$ ARF $<$ \underline{RRAA(A-RTS)} $<$ \underline{RRAA} & \\
    & UDP, hidden terminal, 802.11b &  AARF $<$ SampleRate $<$ ARF $<$ \underline{RRAA(A-RTS)} & \\ 
\hline

\multirow{6}{1cm}{OTA}
     & 5GHz, static, UDP &  Athreos MIMO RA $<$ RRAA $<$ SampleRate $<$ \underline{MiRA} &
     \multirow{6} {1cm} {\cite{Ioannis2010}} \\
    & 5GHz, static, TCP & SampleRate $<$  Athreos MIMO RA $<$ RRAA $<$ \underline{MiRA} & \\
    & 2.4GHz, static, UDP & SampleRate $<$  RRAA $<$  Athreos MIMO RA $<$ \underline{MiRA} & \\
    & UDP, mobility  & SampleRate $<$ Athreos MIMO RA $<$ RRAA $<$ \underline{MiRA} & \\
    & TCP, mobility  & SampleRate $<$ Athreos MIMO RA $<$ RRAA $<$ \underline{MiRA} & \\
    & hidden terminal & Athreos MIMO RA $<$ SampleRate $<$ \underline{MiRA}(\emph{nid}) $<$ RRAA $<$ \underline{MiRA} & \\
\hline

\multirow{5}{1cm}{OTA}
     & static &  Onoe $<$ ARF $<$ CHARM $<$ RRAA $<$ SampleRate $<$ AMRR $<$ \underline{RAM} &
     \multirow{5} {1cm} {\cite{xichen2012}} \\
    & mobile (walking) & Onoe $<$ SampleRate $<$ AMRR $<$ ARF $<$ RRAA $<$ CHARM $<$ \underline{RAM} & \\
    & mobile (driving) & Onoe $<$ AMRR $<$ ARF $<$ SampleRate $<$ RRAA $<$ CHARM $<$ \underline{RAM} & \\
    & mobile, indoor & AMRR $<$ Onoe $<$ ARF $<$ SampleRate $<$ RRAA $<$ CHARM $<$ \underline{RAM} & \\
    & indoor with interference &  Onoe $<$ AMRR  $<$
    ARF $<$ SampleRate $<$ RRAA $<$ CHARM $<$ \underline{RAM} & \\
\hline

\multirow{3}{1cm}{OTA}
     & UDP 802.11a, static & ARF $<$ RRAA $<$ \underline{SGRA} &
     \multirow{3} {1cm} {\cite{SGRA}} \\ 
    & hidden terminal & ARF $<$ RRAA $<$ \underline{SGRA} &\\
    & 2.4GHz interference & RRAA $<$ \underline{SGRA} & \\
\hline 

\multirow{4}{1cm}{OTA}
     & indoor, minor contention &  Onoe $<$ AMRR $<$ SampleRate &
     \multirow{4} {1cm} {\cite{Ancillotti2008}} \\
    & indoor, high contention & AMRR $<$ Onoe $<$ SampleRate & \\
    & outdoor, directional &  Onoe $<$ AMRR $<$ SampleRate & \\
    & outdoor, omni-directional  & AMRR $<$ Onoe $<$ SampleRate &\\ 
\hline

\toprule    
\multicolumn{4}{l}{
\textbf{Note:} SIM: simulation; EMU: emulation; OTA: over-the-air; \underline{Underline}: mechanism proposed in Ref; (\emph{id}) / (\emph{nid}): with/without interferer detect; }\\ 

\label{tab:exp_MAC_layer}

\end{longtable}
\end{landscape}


\subsubsection{Simulation Based Evaluation}

Mythili et al. \cite{Vutukuru2009} evaluate the performance of SoftRate, RRAA, SampleRate, an RBAR-like mechanism and a CHARM-like mechanism using NS3 simulations. The RBAR-like mechanism enables the acknowledgements to carry the SNR feedback and the RTS/CTS mechanism is disabled, while the CHARM-like mechanism uses the average SNR over multiple frames. The topology in the experiment is an N-clients to 1-AP (Access Point) scenario and the traffic is modeled as TCP flows with a frame size of 1400~bytes. The slow fading channel is simulated by the sender moving away from the receiver at a walking speed while the fast fading channel is simulated by the speed of a vehicle. The hidden terminal interference is simulated by the imperfect carrier sense and the carrier sense probability is set to between 0 and 1. In the slow fading scenario, SoftRate shows higher throughput than SampleRate, RRAA, CHARM and RBAR. SampleRate has the lowest throughput in the scenario. In a fast fading channel, SoftRate shows the best throughput and RBAR has the lowest throughput. In the hidden terminal interference scenario, SoftRate also outperforms other mechanisms, and the lowest throughput is achieved by RRAA.

In \cite{BEWARE}, BEWARE, CARA, ARF and ARF-RTS (ARF with RTS turned on) are compared in the varying contending station scenario and the Ricean fading scenario. In the varying contending station scenario, the number of contending stations is varied from 0 to 15 and CARA shows the best performance. In the Ricean fading scenario, the Ricean parameter $K$ and doppler spread $f_m$ are varied. When $K$ increases, the line-of-sight component is stronger and the SNR increases. When $f_m$ increases, the channel condition changes faster. BEWARE achieves the best performance.

In \cite{saadiscc2008}, ARF, CARA and ERA are compared in a static, mobile and contention network. The contention network is simulated with different levels of hidden terminal collisions. ERA outperforms ARF and CARA in all scenarios. 

SIRA+SGRA is compared with SGRA in \cite{Okhwancomletter2015} by varying the walking speed from 0 to 1 m/s. SIRA is an intra-frame algorithm which selects multiple rates for a single aggregate frame to be transmitted. SIRA+SGRA combines the intra-frame rate adaptation, SIRA, and the inter-frame rate control, SGRA. The results reveal that SIRA+SGRA outperforms SGRA significantly in mobile environments.

AARC, Minstrel, RRAA and CARA are evaluated on the NS3 platform in a slow fading scenario \cite{lixingICC2015}. The slow fading channel is simulated by the walking speed of 5 m/s. It shows AARC has the best performance and Minstrel has the worst performance.

In \cite{MutFed}, MutFed, ARF and AARF are compared in a mobile scenario on the OPNET platform. The transmitting node moves away then towards the receiver. MutFed shows the lowest retransmissions, minimal delay and highest throughput.

\subsubsection{Emulation Based Evaluation}
Three rate control mechanisms plus two of their extensions are implemented and evaluated by Camp et. al \cite{Camp2008a} on the WARP platform. The evaluated mechanisms include ARF, RRAA, RRAA with the A-RTS extension, RBAR, and RBAR with the OAR extension. OAR \cite{Sadeghi2002} is an extension which enables rate adaption mechanisms to send back-to-back frames. The authors use a Spirent Communication Channel Emulator (SR5500) to emulate various channel conditions. They vary the coherence time from 100~$\mu$s to 100~ms on a single channel with a high average RSS of -40~dBm to evaluate the throughput. The coherence time refers to the time interval over which the channel condition is sufficiently constant to decode the received symbols. For a long coherence time (bigger than 100 ms), all mechanisms converge to similar performance except RBAR. In a short coherence time (less than 1 ms), the highest throughput is achieved by ARF and RBAR with OAR extension.

CHARM, Onoe, SampleRate and AMRR are also evaluated in \cite{Judd2007}\cite{Judd2008} in a controllable environment to compare their performance  in a mobile environment and to study the impact of hidden terminals. In the mobile environment with the speed from 0.5 m/s to 2 m/s, CHARM performs the best and Onoe has the lowest throughput. In the hidden terminal environment, CHARM and SampleRate outperform Onoe and AMRR rate control mechanisms.

In \cite{YIN2010}, AMRR, Onoe, SampleRate and Minstrel are compared in a controllable environment for three scenarios. The first evaluates performance under fixed channel conditions with different SNR. The second compares the performance under a general interference environment. The impact of interference duration, interference interval and interference strength are evaluated. The last scenario is to evaluate the effect of hidden terminal problems. In fixed channel conditions, SampleRate shows the best performance.  In a general interference environment, SampleRate, AMRR and Onoe cannot select the optimal transmission rate when there is strong interference, thus achieves the lowest throughput. In the hidden terminal scenario, the impact of hidden terminal traffic load and heterogeneous link quality are evaluated. Minstrel shows the best performance, since it does not behave too aggressively on the good link, nor experiences starvation on the bad link.

In \cite{Halperinsigcomm2010}, ESNR, SampleRate and SoftRate are compared in SISO (Single Input Single Output) mobile environments. ESNR performs slightly better than SampleRate and SampleRate performs a bit better than SoftRate in both human-speed and fast mobile environments. 

Minstrel-rts, Minstrel, PIDE and PID are evaluated in \cite{mac80211ratecontrol2015} for the hidden terminal scenario.  The hidden terminal traffic load is varied from 1 Mbps to 17 Mbps. Minstrel-rts displays the best performance and Minstrel is the second. PID has the lowest throughput.

PIDE, PID and Minstrel are evaluated on a controllable platform in \cite{MSWiM2011}. PIDE shows the best performance in static, linear increase/decrease and suddenly changing channel conditions.

In \cite{weilcn2012}, performance comparison is conducted for RCELC and Minstrel
in static, linearly increasing/decreasing, and sudden changing channel conditions. The results confirm RCELC outperforms Minstrel.

\subsubsection{Over-the-air Evaluation}
ARF, AARF, Onoe and SampleRate are evaluated in \cite{Bicket2005} on a 45-node indoor test-bed and a 38-node outdoor test-bed. Packets used are 1500~byte UDP packets. In \cite{Bicket2005}, the authors found that ARF achieves the lowest throughput in both outdoor and indoor networks. Onoe achieves a very close performance as the best fixed rate in indoor experiments but it performs poorly in low quality links in the 802.11a and 802.11b outdoor networks. SampleRate achieves the best throughput on both the indoor and outdoor networks even on lossy links.

Wong, et al. \cite{Wong2006} evaluated RRAA, SampleRate, ARF and AARF in the IEEE 802.11a/b networks with various settings, such as static/mobile clients, with/without hidden stations. In a scenario with stationary clients, the results for four clients in the IEEE 802.11a network show that RRAA has the best performance and SampleRate has better performance than ARF and AARF in the IEEE 802.11b static networks. In a mobile-client scenario with UDP packets, SampleRate has lower throughput than RRAA, ARF and AARF. In the hidden terminal scenario, RRAA with the A-RTS mechanism has the best performance. The experiments in \cite{Wong2006} are conducted in a real environment with limited nodes and the result is subject to various client positions. It is not clear how the hidden terminal scenario is created in the real environment and it is obvious that it will be difficult to achieve a repeatable evaluation environment.

SampleRate, AMRR and Onoe are evaluated in \cite{Ancillotti2008} in both indoor and outdoor environments. The indoor environment has 12 nodes on one floor, while the outdoor environment has five wireless routers deployed on the rooftops of three buildings. Experiments were carried out in the 802.11g mode using UDP traffic. In good  quality channels in the indoor environment where the highest throughput is achieved by 54Mbps, AMRR and SampleRate show better performance than Onoe. A high contention channel condition is formed in the indoor environment by enabling eleven senders to transmit frames concurrently. In the scenario, SampleRate sends only 7\% of frames at the optimal rate. AMRR and Onoe perform even worse than SampleRate. In low-mid medium contention channel condition, SampleRate also shows better performance than AMRR and Onoe. However, it is not clear how to determine the degree of contention in the environment. In the outdoor environment, SampleRate also shows better performance than AMRR and Onoe. 

In \cite{dongxiaicc2013}, the Minstrel rate control mechanism is compared with fix rates in the 802.11g networks. Experiment results show that Minstrel performs well in static channel conditions but fails to select the optimal rates in some cases of the dynamic channel conditions. Specifically, when the channel degrades from good to bad, Minstrel does not perform as well as in the channel condition when the channel improves from bad to good.

MiRA, RRAA, SampleRate and Atheros MIMO rate adaptation  \cite{Ioannis2010} are compared in static, mobile and hidden terminal environments. Atheros MIMO rate adaptation is based on Minstrel but it modifies specific features to cater for the MIMO environments. The results show that MiRA has better performance than other rate control mechanisms in all scenarios. 

SampleRate and FARA are evaluated in \cite{Rahul2009} for 17 different locations using 20 MHz and 100 MHz channel and FARA performs better than SampleRate for all locations.

HA-RRAA, SampleRate, RRAA and ARF are compared in different scenarios in \cite{Ioannis2013}. These scenarios include UDP and TCP flows, static and mobile settings, hidden terminals, using 2.4 GHz or 5 GHz bands. HA-RRAA has the best throughput in all scenarios except the mobile UDP environment using the 802.11a channel.

In \cite{xichen2012}, Onoe, ARF, CHARM, RRAA, SampleRate, AMRR and RAM are evaluated in static, mobile at walking or vehicle speed, in-door mobile and indoor with interference environments. RAM displays the best performance in all scenarios. Onoe almost achieves the worst for all experiments except the indoor mobile scenario.

SGRA, RRAA and ARF are compared \cite{SGRA} in the static 802.11 UDP environment, hidden terminal environment and the environment with bluetooth interference. SGRA outperforms other mechanisms in all scenarios.

%% file: critiques.tex
\section{Lessons learned}
\label{sec: lessonlearned}

In this section, we conclude our study of rate control mechanisms at the MAC-layer and share the following important lesson learned with researchers developing new rate control mechanisms. 

\subsection{Decreasing transmission rate upon severe frame loss is not a good reaction if collisions exist.}

It is generally recognized by most existing mechanisms \cite{PID} \cite{Karmerman1997}  \cite{Lacage2004a} \cite{Onoe}  that the transmission rate must be decreased if intense frame losses occur. The motivation for this reaction is that the underlying problem behind significant frame losses is the deteriorating link condition between the sender and receiver, and that the current rate must be decreased to cater for the worsening link. 

The above rule holds for most cases where the link condition between the communicating pair degrades, but breaks if collision occurs. Both hidden terminal and congestion collisions can cause significant frame losses that are independent of the channel quality. To decrease the transmission rate under such conditions will not solve the problem but make it worse, as lower rates have longer transmission times and the increased transmission time for each frame aggravates collision.

Therefore, the guideline of decreasing the rate upon severe frame losses does not hold in collision loss cases. The rate adaptation mechanism needs to differentiate various losses and react accordingly.
 
\subsection{Increasing transmission rate only because of low frame loss may cause rate oscillation problem.}

Most of the existing mechanisms \cite{PID} \cite{Wong2006}  suggest that the transmission rate must be increased if the frame loss ratio is low. The underlying conjecture is that the link condition supports the current rate so well that it should support the next higher rate. 

This conjecture holds true for most cases, but fails in some cases where the current rate is already the best, yet with a low loss ratio. From Fig. \ref{fig:fixrate_rss}, the horizontal curve of a rate stands for its maximum achievable throughput, with the frame loss ratio equal to zero. Its next higher rate's curve intersects with the rate's curve. The horizontal curve is divided into two parts by the intersection point. For example, the intersection point of 48 Mbit/s and 36 Mbit/s is path loss 77dB. On any point of the left horizontal curve (before 77dB), the rate must be increased because the throughput of the next higher rate is higher. On the right side of the horizontal curve (after 77 dB but before 81 dB), although the rate has a zero frame loss ratio, the rate cannot be increased because the throughput of the next higher rate is lower. For right side points, the loss ratio of the next higher rate is high. If the rate is increased, the high loss ratio will force the rate to decrease again. This causes the rate oscillation problem in rate selections, as discussed in PID \cite{PID}. 

Therefore the proposed higher rate should be checked whether the new rate is able to provide better throughput before switching to it. This is one of the improvements we incorporated into PIDE in \cite{MSWiM2011}.

\subsection{Using only success or failure information of probing frames may mislead rate adaptation decisions.}

The probing approach is utilized in several rate control mechanisms \cite{minstrel} \cite{Lacage2004a} \cite{Bicket2005} for rate adaptation. Probing frames are those frames that are transmitted at rates other than the current rate. Without probing, a rate control mechanism's knowledge is limited to the current rate's performance and it has to make a rate decrease/increase decision just based on the limited knowledge.  As discussed in the last subsection, in Fig. \ref{fig:fixrate_rss}, the horizontal region of a rate's throughput curve is divided into two parts by the intersection point with its next higher rate. Throughout the horizontal region, the current rate performs very well but the next higher rate's performance is very different among the two parts: good on the left but very bad on the right. This means based on the current rate performance, we cannot predict higher rates' performance. Therefore, probing is necessary.

However, using a binary probing result, i.e., the probing frame is successful or not (e.g. AARF \cite{Lacage2004a} ), is not an adequate method. For some point in Fig. \ref{fig:fixrate_rss}, it can have near-perfect transmission at 12 Mbps, but almost 40\% frame loss at 18 Mbps. This means, a probing frame at 18 Mbps has a probability of 60\% to get through and the rate will be likely increased. However, 18 Mbps with 40\% frame loss achieves lower throughput than 12 Mbps with zero frame loss. Therefore, binary probing results can mislead the rate control mechanism to a wrong decision.

The throughput is a direct performance metric to reflect the most common goal in terms of network optimization, i.e., maximizing the aggregate throughput. It is probably better than the binary transmission status of the probing frames. An appropriate way for probing could estimate the throughput metric \cite{MSWiM2011} and make a rate decision based on throughput comparisons. 

\subsection{Using the physical layer metric for rate adaptation decision may cause a standard compliance problem or is based on unrealistic assumptions.}

A number of proposals use  physical layer metrics including SNR and BER to estimate the link quality and make rate adaptation decisions. However, BER and SNR can not be measured locally at the senders, without making a symmetric link assumption. According to where the transmission rate is selected, SNR based mechanisms are classified into sender controlled, e.g., CHARM \cite{Judd2007}, SGRA \cite{SGRA} and LA \cite{LA},  and receiver controlled mechanisms, e.g., MutFed \cite{MutFed} and RAM \cite{xichen2012}. For sender controlled mechanisms, the sender estimates the SNR at the receiver side. These mechanisms assume link symmetry between the sender and receiver, i.e., they assume the path loss of sender-to-receiver and receiver-to-sender is the same and both the sender and receiver experience the same level of interference and noise. This assumption does not hold true in wireless environments. For receiver controlled mechanisms, SNR is measured at the receiver and the rate is selected by the receiver. Then the rate is notified to the sender by transmitting ACK using the selected rate, which causes the standard compliance problem. Moreover, SNR is measured during the reception of the preamble. The later part of the frame may experience fading and interference, which is not captured by the SNR at the preamble. In addition, SNR may not be accurate because of hardware calibration  and interfering transmissions \cite{xichen2012}.

\subsection{Using simple heuristics to address hidden terminal problem cannot maximize throughput gain.}

Some rate control mechanisms use simple heuristics to address the hidden terminal problem. In CARA \cite{CARA}, RTS is not turned on for the first retransmission attempt since the first transmission failure may be caused by collisions. If the first retransmission fails again, the loss is attributed to channel errors as RTS should have already mitigated the collisions. Therefore a lower rate will be applied. RRAA \cite{Wong2006} uses A-RTS to solve the hidden terminal problem. A window for monitoring the performance of RTS/CTS is updated in RRAA. Its initial value is zero, which means disabling RTS. If a frame is lost without RTS, the window is increased by one, as the frame may have collided. If a frame is lost with RTS, or succeeds without RTS, the window is halved as RTS does not contribute in both cases. 

In CARA, when long collisions exist, the first attempt of a frame will always fail so it still has at least 50\% losses, meaning 50\% throughput loss. In RRAA, the halving approach seems to be aggressive as well. Both approaches are based on simple heuristics and do not make a qualitative analysis on whether turning on RTS can  enhance performance. 

A promising approach proposed in Minstrel-rts \cite{mac80211ratecontrol2015} could solve the hidden terminal problem better. It makes a comparison between the throughput achieved with RTS and without RTS, and then decides whether to turn on RTS and for how long. When calculating the throughput achieved by RTS, the overhead of RTS/CTS frames and benefits (less losses) are also considered, so that the throughput gain can be maximized. 

\textbf{Newer networks such as IEEE 802.11ac offer a feature, called \emph{channel bonding}, to bind multiple 20~MHz channels together for higher bandwidth. In some cases, not all 20~MHz channels are free for transmission. IEEE 802.11ac extended the RTS and CTS to add \emph{bandwidth signaling} \cite{Matthew2015}. This new feature makes use of RTS/CTS to detect/avoid hidden terminals and to identify which 20~MHz channels are free for channel bonding. To request a channel, the initiator sends  separate RTS frames on the four primary 20~MHz channels. The receiver only responds with CTS on the 20~MHz channels that are free from interference. However, this approach still incurs the RTS/CTS overhead.}

\subsection{Considering surrounding traffic load causes rate adaptation not to converge to the best rate.}

Some mechanisms \cite {BEWARE} suggest that the background traffic and the traffic load must be considered. The underlying conjecture is that it prolongs the actual transmission time for delivering a frame.

BEWARE \cite {BEWARE} makes rate adaptation based on the transmission time. When calculating the performance metric, it adds into calculation the backoff time caused by a neighboring node's transmission. Therefore, varying traffic load in the background could cause rate change. For example, suddenly increased traffic load increases the transmission time of the current rate, which may result in rate decreases. The dependency of rate adaptation decision on background traffic makes rate control mechanism load aware and makes it difficult to converge to the best rate if the traffic load is continuously changing. 

Should a rate control mechanism makes use of network load in rate adaptation, it should be aware of the overhead and negative impacts involved.

\subsection{Assuming a unique FLR threshold for all rates cannot maximize the throughput.}

Some rate control mechanisms assume a unique FLR threshold (PID-14\%, AMRR-33\% and Onoe-50\%) to decrease the transmission rate. This is based on the assumption that the current rate with FLR that is above the threshold achieves lower throughput than its next lower rate.

Fig. \ref{fig:fixrate_rss} shows the throughput achieved by all fixed rate in the IEEE 802.11a mode. From Fig. \ref{fig:fixrate_rss} we can see that the absolute maximum throughput for a given path loss is on the envelope that connects the maximum throughput points of all the fixed rates. Hence the optimal maximum FLR of each fixed rate are the crossing points when a higher rate switches to a lower rate to maintain the highest throughput. When we compared the FLR of these throughput points in Fig. \ref{fig:fixrate_rss}, we found that they are all different values. Therefore, using a fixed target FLR  for all rates fails to achieve the maximum throughput. 

When calculating the FLR threshold to decrease the rate, FLR based mechanisms should accurately compute the FLR of the crossing points of adjacent rates' throughput in order to ensure maximum throughput. In \cite{MSWiM2011}, Yin et al. conducted detailed experiments to verify that there is no single best FLR threshold which can help to achieve the maximum throughput at all time.

%% file: open_issue.tex
\section{Future research}
\label{sec:newDevelopment}

In the last two decades, many rate control mechanisms have been proposed for 802.11 networks. However, rate control mechanisms are also required in other wireless networks. Other wireless networks, e.g., body area networks~\cite{linchihan2014} and vehicular networks~\cite{yuanieeecl2014}, have some features in common with 802.11 networks, such as  wireless shared medium, interference, and fading characteristics. However they also have some specific features and therefore  rate control mechanisms designed for 802.11 networks may not be suitable for other networks. For example, rate control mechanisms for 802.11 networks operate in a unicast mode. Feedback, e.g., ACKs, are utilized to provide channel quality information to the sender. Vehicular networks use a broadcast mode, in which a sender will not receive ACKs but others' transmissions instead. The challenge for vehicle networks is how to estimate the channel condition without ACKs. In body area networks, a number of sensors are worn/implanted to exchange information and monitor data. The channel characteristics are highly dynamic as a human body  moves with gestures, posture change or human mobility~\cite{Maskooki:2013aa}. This is different from traditional Wi-Fi networks, where nodes are relatively stationary. The suitability of rate control mechanisms from 802.11 networks for other networks should be investigated before being applied to these networks.

\subsection{New opportunities}
\subsubsection{Features in  new standard}
The new feature of 802.11n compared to 802.11abg is the 
Single User Multiple-Input Multiple-Output technology (SU-MIMO). Two operational modes are supported by the 802.11n standard. One is the diversity-oriented single-stream (SS) mode. The other is the spatial multiplexing-driven, double-stream (DS) mode. With the two modes, 802.11n allows many more rate options than the 802.11abg mode, ranging from 6.5~Mbps to 600~Mbps. As the rate increases, loss does not monotonically grow with rates in different modes~\cite{Ioannis2010}. Therefore, the existing rate control mechanisms designed for 802.11abg cannot be applied to the 802.11n networks. Although a few  rate adaptation algorithms, e.g., MiRA~\cite{Pefkianakiston2013}\cite{Ioannis2010}, are designed for the 802.11n networks, the characteristics of the 802.11n technology have not been fully explored. For example, frame aggregation is an important MAC enhancement in the 802.11n networks. One type of frame aggregation methods is A-MPDU (Aggregated MAC Protocol Data Unit), which requires a long frame duration, e.g, the maximum of 10ms. This exceeds the coherence time in mobile environments~\cite{Camp2008}. Therefore a rate control mechanism should be designed to adapt rates within an individual frame~\cite{Okhwancomletter2015}. Even for the 802.11abg mode, recent work, e.g. \cite{Hyunjoongicc2014},  shows the necessity to make rate adaptation during a frame transmission in fast fading environments. Other rate control mechanisms designed for the 802.11n networks, including InFRA, are based on effective SNR. To calculate the metric, the receiver needs to compute BER and then map BER to effective SNR. BER is not directly a reflection of the throughput so it cannot guarantee to achieve maximum throughput. 

\textbf{In 802.11ac networks, multiple single-antenna clients communicate concurrently with a multi-antenna AP~\cite{weiliang2014} to form a Multi-user (MU) MIMO transmissions. The rate decisions made by clients are not independent and they interact with each other, which is different from the existing 802.11 networks. This poses a new challenge for a rate control design.  802.11ac uses a compressed form of CSI (Channel State Information) to reduce the overhead of sending full CSI feedbacks. In this case, 802.11ac APs can only depend on MAC-layer feedbacks, e.g. packet-error-rate (PER) and PHY rate statistics (MCS) to decide MU-MIMO groups~\cite{Surmobicom2016}. Only users that have similar profiles (e.g. same channel bandwidth and throughput) should be classified into the same group. This is to overcome the efficiency problem caused by heterogeneous bandwidth users. However, the profile parameters depend on the rate control metric. Therefore, the rate control mechanism should be combined with group user decision to enhance network throughput in 802.11ac networks.}

\textbf{The IEEE 802.11ad technology doubles the bandwidth provided by the 802.11ac technology, but sacrifices it for a very short transmission range of a few meters. This means the transmission rates are sensitive to the distance between the two communicating nodes. Mobility causes the distance change very frequently so that rate control mechanisms should be very responsive to mobility. In this case, methods such as using a shorter rate adaption period and putting more weight on fresh metric measurements and employing per-packet rate selection can be adopted to improve performance.}

\begin{figure}[t]
\centering
\includegraphics[width=\linewidth]{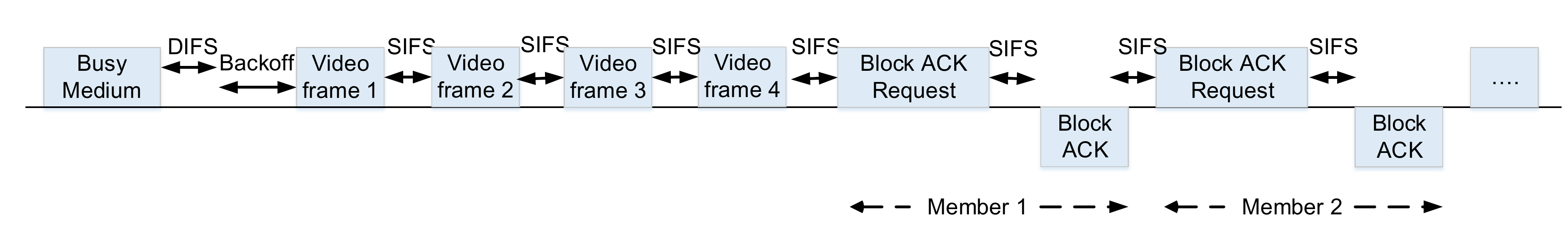}
\caption{Block ACK mechanism in 802.11aa}
\label{fig:802.11aablockack}
\end{figure}

\textbf{In the 802.11e standard, multicast frames are transmitted at one of the 'Basic Service Set' rates. However, basic rates are very low and due to the anomaly of link quality among group members, sending rates at one of those rates may decrease the overall performance. In addition, retransmission of lost multicast frames is disabled, so the robustness is not ensured. The 802.11aa standard \cite{802.11aa-amend} aims to enhance the efficiency and robustness of multicast video streaming in WLANs. The Block ACK mechanism, as shown in Fig.~\ref{fig:802.11aablockack}, is utilized to ensure reliability. The rate control mechanism is recommended to adapt all available transmission rates. However, the specification is not standardized~\cite{Kaouther2017}. The block ACK contains frame loss information so that it can be leveraged to calculate the link quality metrics, e.g. frame delivery ratio and throughput, to select an appropriate rate. One difference from rate control in 802.11 unicast networks is that the sender should select an appropriate rate for all multicast links instead of just one link. Therefore, a metric that considers all counted links should be designed. In \cite{Kaouther2017}, Kaouther et al. present a mechanism based on the FLR metric. However, they fail to address the rate oscillation problem inherited from FLR-based mechanisms. In \cite{salvador2013first}, Pablo et al. present a first 802.11aa implementation  on commodity cards and make it publicly available; this can be utilized by researchers to develop and evaluate rate control mechanisms for 802.11aa in real environments.}

\textbf{The IEEE 802.11ah WLAN protocol allows a longer transmission range between an access point (AP) and clients, up to multiple kilometers~\cite{StefanICC2012}. Therefore, it is widely used in many fields including health care, Internet of Things, non-intrusive remote sensing and UAVs. Most applications are deployed in the outdoor environments with more node density, mobility and interference. Therefore, rate control mechanisms should combine with WLAN configurations at the PHY layer and the MAC layer \cite{Stefan2015} to tackle challenges from contention loss, mobility loss and interference loss that are more severe than in an indoor environment.}

\textbf{The 802.11ax technology is to replace the 802.11n and 802.11ac technology in the long run. In  802.11ax networks, the transmission power and CCA levels can be adaptively changed~\cite{BellaltaIWC2016}. Increasing the CCA levels can reduce the influence area and increase the chance to transmit, hence improving the throughput. Decreasing the transmission power reduces the influence area, which improves the spatial reuse. However, the number of packet errors may increase and lower transmission rate may be selected. Therefore, rate control mechanisms should consider the impact of transmission power and CCA level changes.}

\subsubsection{limitations in rate control}
\begin{figure*}[t]
\centering
\subfigure[ACK spoofing]{
   \includegraphics[width=0.35\linewidth] {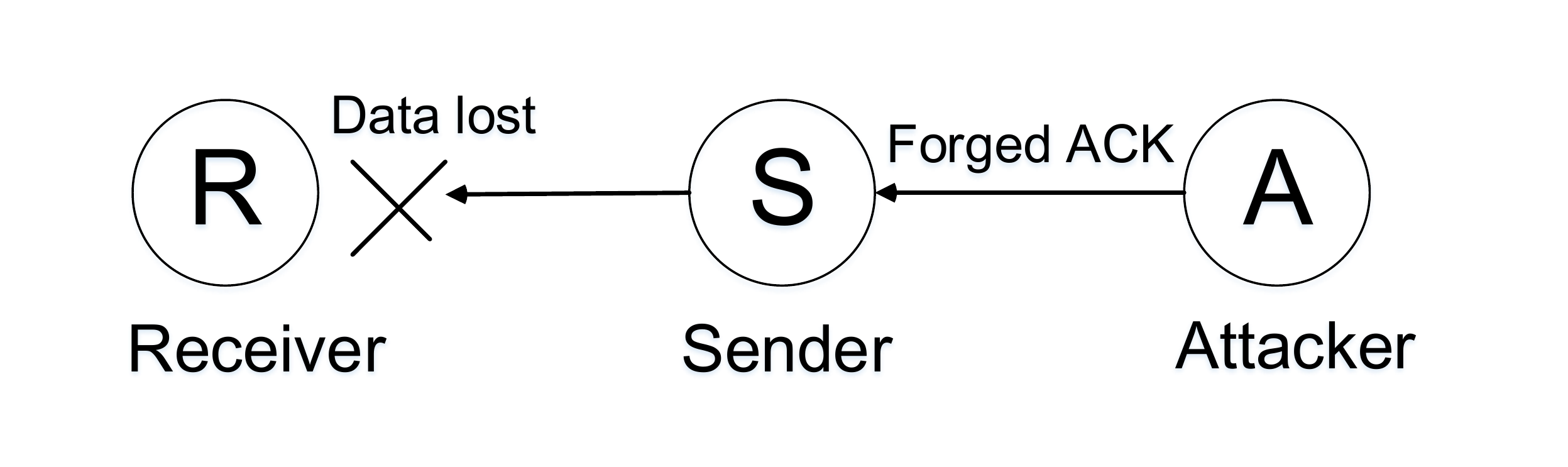}
   \label{fig:ACKspoofing}
 }
 \subfigure[Probe spoofing]{
   \includegraphics[width=0.35\linewidth] {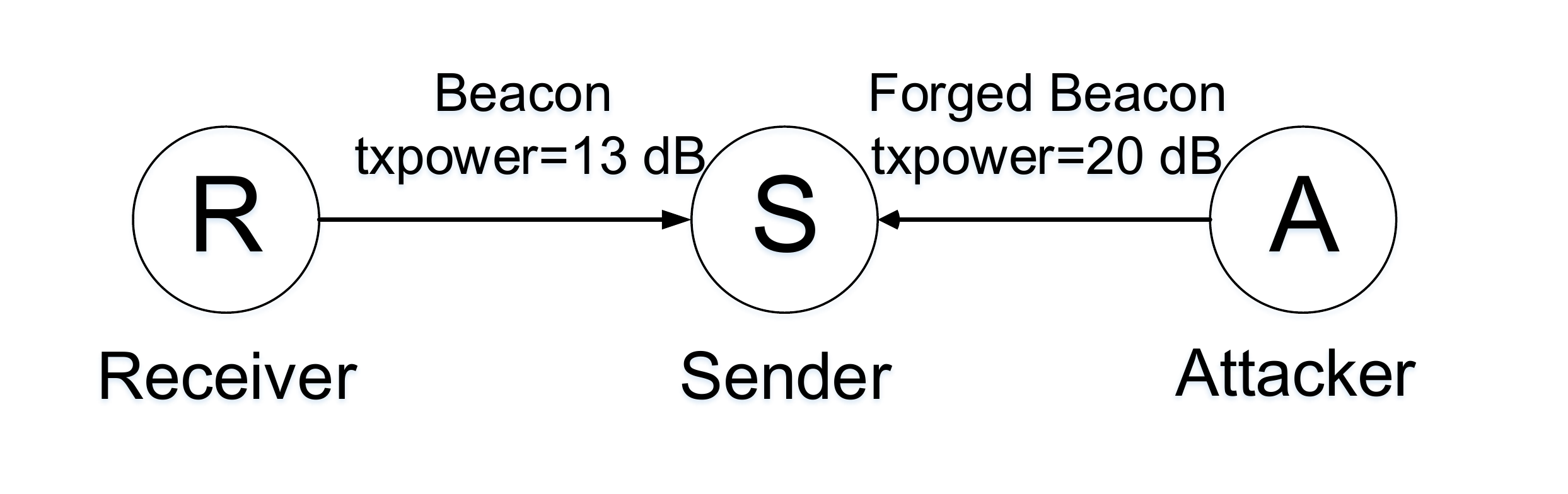}
   \label{fig:probeattack}
 }
  \subfigure[Jamming]{
   \includegraphics[width=0.35\linewidth] {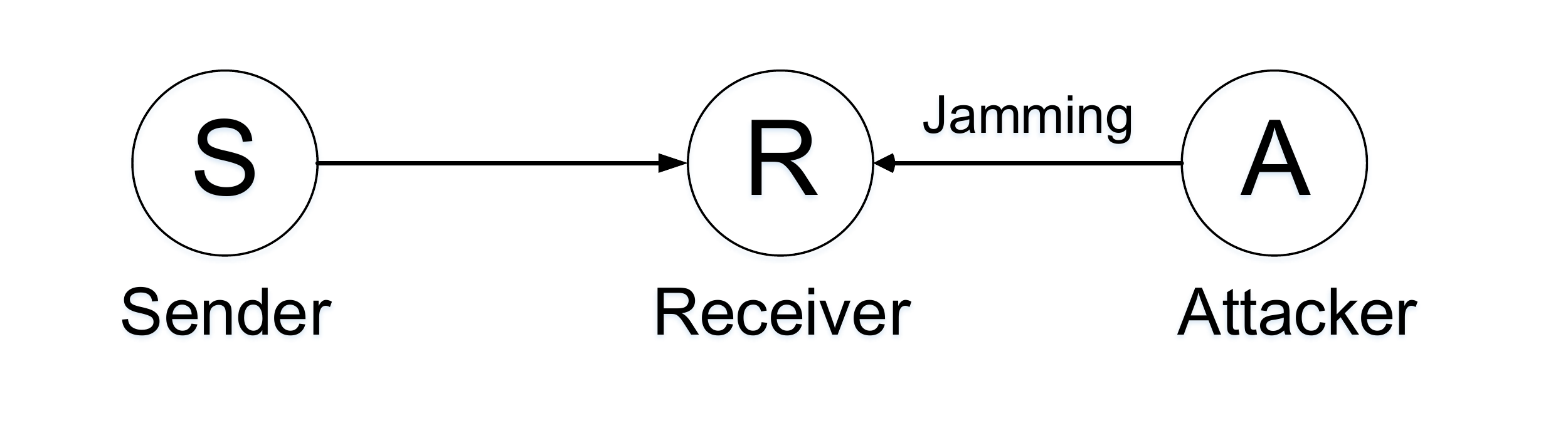}
   \label{fig:jamming}
 }
\caption{Attacks against rate control mechanisms}
\label{fig:attacks}
\end{figure*}

\textbf{The sender in ACK-based rate control mechanisms calculates the link quality metric, such as CTR (e.g. ARF and AARF), FLR (e.g.
PID and RRAA), transmission time (e.g. SampleRate) and throughput (e.g. Minstrel and RCELC) based on acknowledgements. They make rate adaptations according to the monitored metrics. If an attacker forged ACKs to acknowledge those lost frames, as shown in Fig.~\ref{fig:ACKspoofing}, this can interfere with the link quality metric calculation, misleading the rate control mechanism to believe the channel condition is very good and select a higher rate that is not supported. In some SNR-based mechanisms, e.g. CHARM, the sender calculates the SNR based on the transmission power advertised in beacons or probes. Therefore, an attacker can fabricate beacons or probes with a fake transmission power, as shown in Fig. \ref{fig:probeattack}, which affects rate selection, misleading the sender to choose an inappropriate rate. Therefore, rate control should be combined with a mechanism that can verify the ACK or probe authenticity.}

\textbf{In BER-based mechanisms, the BER is utilized for rate adaptation. An attacker can jam the channel, as shown in Fig. \ref{fig:jamming}, to increase the BER to mislead the sender to choose a lower rate. Therefore, BER-based rate control should be able to differentiate jamming errors and channel errors. }

\begin{figure}[t]
\centering
\includegraphics[width=0.25\linewidth]{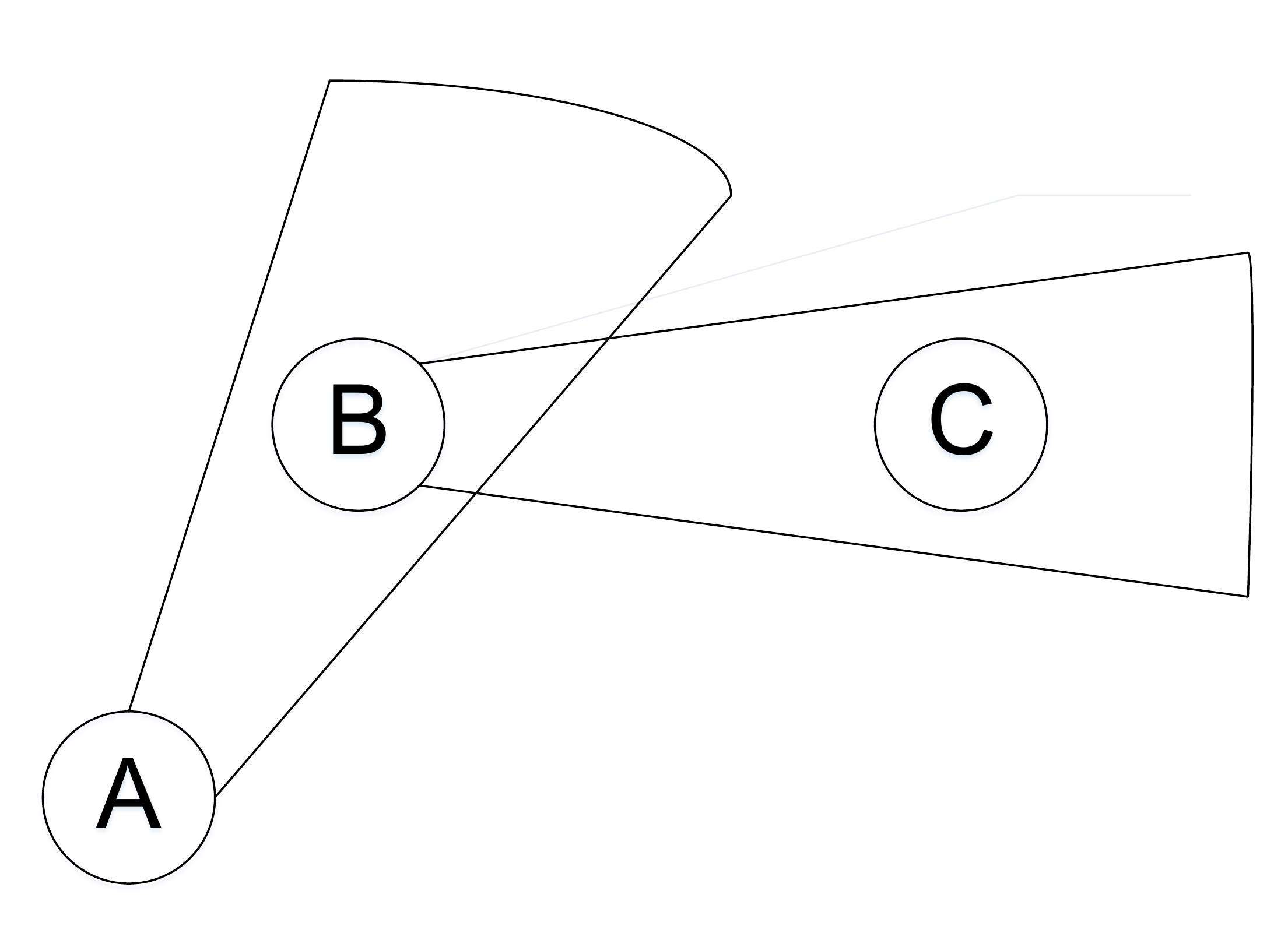}
\caption{Deafness problem for 802.11 networks.}
\label{fig:deafness}
\end{figure}

\textbf{When directional antennas are utilized in 802.11 networks, deafness \cite{Romittech20} may pose a problem for rate control mechanisms, as shown in Fig. \ref{fig:deafness}. Node A may try to send a frame to node B which is transmitting a frame to node C. Because A cannot hear B's directional transmission and B cannot respond with an ACK to A, A concludes that the frame is lost. This can have a big impact on A's rate metric calculation, which leads A to select a low rate. RTS/CTS can solve this problem. When hearing B's RTS frame, A knows B is transmitting. However, when RTS/CTS is disabled, the rate control mechanism should figure out the deafness situation without decreasing the rate. 
}

\subsubsection{The mac80211 framework}

The mac80211 framework is a wireless infrastructure designed for the new Linux Kernel. The framework incorporates common features, e.g., rate control, to simplify a wireless driver design and implementation \cite{mac80211ratecontrol2015}. This framework can support a large amount of research work related to rate control. First, new mechanisms can be designed, implemented and evaluated. Underlying wireless drivers report to the framework the attempted rates, respective retry counts, and the signal strength for the received ACKs after delivering a data frame. Therefore, mechanisms based on FLR, SNR, throughput, transmission time, etc. can be implemented on this framework with ease. Second, most of the existing rate control mechanisms make rate adaptations based on the rate adaptation period (RAP). For example, 10s for SampleRate, 100ms for Minstrel and 1s for Onoe. However, the effect of RAP has not been explored. A large RAP indicates less computation overhead but also less adaptivity to a fast changing channel. Using the framework, a comprehensive study could be carried out to discover the underlying impact for various RAP settings. Third, a cross-layer design is supported by the framework. For example, the 802.11s mesh protocol is built on the framework to enable ad hoc communications among clients so the interactions between rate control mechanisms and the mesh routing protocol can be supported. Some link metrics that are calculated by the rate control can be shared by the mesh routing protocol. Performance enhancement may be achieved through a cross-layer optimisation. Fourth, the multi-rate feature can  cause a performance anomaly problem \cite{Rosario2007} where performance degradation will be suffered if one client has a very bad quality link to the access point (AP). The mac80211 framework can be utilised as a platform to study and improve the fairness of air channel time in a multi-rate wireless environments \cite{Bhanage2010}.
Fifth, as one of the representative techniques in 802.11ac and the
upcoming 802.11ax, Multi-User MIMO promises significant throughput gains by supporting multiple concurrent data streams to a group of
users. However, this design introduces challenges in selecting the best-throughput for MU-MIMO groups due to, 
a) lacking of full CSI feedback from the user, which is commonly used for  MU-MIMO grouping. 
b)  hardware capability and external interferences, which results in heterogeneous bandwidth users that not all users in the group can support the same channel bandwidth. 
c) limited onboard resources on APs, which limits the complex mathematical and memory intensive
operations are not portable to such platforms~\cite{sur2016practical}. 
With these challenges in mind, new approaches on rate adaption have to study to fully utilize the benefits of IEEE 802.11ax.

\subsection{New applications}
\subsubsection{Increasing contention}
The Internet of Things (IoT) \cite{atzori2010IoT} has gained large popularity nowadays due to its  vision of anytime, anywhere and anymedia, which motivates the development of communication technologies. Sensor networks play an important role in IoT and they help track the target's location, temperature, movement, etc, bridging the gap between the real world and digital world.  In a sensor network, nodes communicate in a wireless multi-hop fashion. Rate control mechanisms in sensor networks need to address the scalability and contention problem since the number of nodes in a sensor network will be very high. Power efficiency problem for the MAC layer rate control mechanism is also a concern in sensor networks, as sensor nodes are power limited. In addition, the largest frame size is usually 102~octets, which is very small.  This means that the ratio of communication overhead (e.g. SIFS and DIFS) to payload transmission time is much higher. Frame aggregation approach may be used in MAC rate control mechanisms to enhance performance. In addition, sensor nodes spend a large amount of time in a sleeping mode, how to select the transmission rate after waking up is a research question. Cars, trains and buses are equipped with sensors to provide better navigation and avoid car collisions. The research question  for rate control mechanisms is how  to minimize transmission time and enhance timeliness and provide high mobility support especially in the high-way system.

\subsubsection{High mobility}
Mobile Cloud Computing (MCC) \cite{dinh2013survey} has emerged as a new technology to move computing and storage from mobile devices to the cloud with powerful and centralized computing platforms.  Therefore, services provided by the cloud are accessed via the wireless communication technology. The mobile devices then are confronted with many challenges such as limited battery life and bandwidth, channel dynamics caused by mobility and interferences. So the design goal of rate control mechanisms may vary and they should address the power efficiency problem or the robustness issue in face of mobility and interference. In addition, various technologies, e.g., WCDMA and WLAN, are used for communication in MCC. These technologies are wireless but have different features, therefore rate control mechanisms for those technologies  need to be investigated separately in order to enhance the aggregate performance.

Unmanned Aerial Vehicles (UAVs) \cite{wu2012cross} plays a key role in modern military battlefields, conducting surveillance and reconnaissance. Communication is enabled between the UAVs and the ground control station for command and control, as well as transmitting captured information including image and video files. The Long-Term Evolution (LTE) technology is usually utilized in UAVs for communication, providing high speed and low latency compared to other technologies including Joint Tactical Radio System (JTRS). Besides high volume of transmitted data and its timeliness, UAVs are highly mobile and energy limited. MAC rate control mechanisms for the LTE technology used in UAVs have gained little attention so far. A good MAC layer rate control mechanism  for this kind of network needs to improve the throughput, reduce the latency, deal with mobility and power efficiency problems. In addition, when applied in military applications, UAVs become the targets for jamming attacks. Orakcal et al. \cite{cankutpv2014} demonstrate that several rate control mechanisms, e.g., SampleRate and ARF, are vulnerable to the jamming attack. Jamming attacks increase the FLR, so some mechanisms may decrease the rate. Therefore, in UAVs, a robust mechanism should differentiate frame losses from jamming, mobility and channel errors.

\subsubsection{Energy awareness}
Traditionally, rate adaptation mechanisms have been designed to maximize network goodput. However, as the battery-powered devices, including smartphones, drones and other smart IoT devices (lightbulbs, doorlocks), are becoming dominated in the wireless networks. There are increasing needs to design rate adaptation algorithms that are energy-efficient. The measurements presented in \cite{lichiyumobicom12} shows that an 802.11n $3\times3$ MIMO receiver consumes about twice the power of 802.11a during an active transmission, and 1.5 times power when idle. 
The authors in~\cite{li2016energy, lichiyumobicom12} claim that most existing solutions are practical to ensure high goodput but not energy savings. 
Their results show ARA~\cite{CARA} and MiRA~\cite{Pefkianakiston2013} incur 
per-bit energy waste at a network interface card as large as 54.5\% and 52.9\%, respectively. 
The factor of power consumption brings the designing of rate adaptation algorithm into a new era which can benefit power-constrained devices. 
In the aspect of the sender, to transmit a data frame, a lower transmission rate consumes more energy than a higher rate due to longer transmission time. However, this is not the case for software defined radio (SDR), in which power consumption not only involves expenditure on frame transmission, but also on the processing of the micro-processors running the SDR software. Low transmission rate reduces the speed and voltage of the micro-processor, hence consuming low energy. Recent research discovers that the effect of running the SDR software on micro-processors is more significant than transmission time if power consumption is concerned \cite{jung2013joint}. Therefore, if the design goal for rate control is power efficiency, rate control mechanisms could consider using lower transmission rates when channel utilization is low.

Also, one assumption that the optimality in throughput means optimality in power consumption in the 802.11 networks has been put in question recently \cite{lichiyumobicom12} \cite{Khanmobihoc2013}. Ucar et al. \cite{UcarMSWiM2016} confirm that the throughput and power efficiency cannot be simultaneously optimized. However, the discovery is based on a theoretical analysis and a practical experimental study is still lacking. If energy efficiency is a concern in the 802.11 networks, the performance metric should be redesigned for existing throughput-oriented rate control mechanisms to maximize power efficiency. However, with a limited knowledge of the practical system, power consumption metrics may be subject to many factors including CSMA backoff, frame losses and the transmission rate, and are hard to be measured accurately for individual frame transmissions. An active learning approach based on machine learning \cite{euhannaOC2016} can be utilized to obtain feedback from the environments to achieve optimal power and rate control.

Cognitive radio technology is capable of learning the status of a frequency band in the environment. Therefore, it is able to switch to a free frequency band when it senses that the currently used frequency is busy. It is observed in \cite{Rahul2009} that the link quality varies according to different frequency bands, exhibited by SNR values. Hence, although the free frequency band experiences less interference than the busy one, the link quality is not necessarily higher and a higher rate is not always supported. A blind switching to a free channel band may cause a rate decrease and hence throughput degradation. Rate control mechanisms in cognitive radio networks should take into account the interference and link quality of a frequency band and then make appropriate frequency and rate selection. On one hand, interference, due to exclusive frequency sharing, determines what  percentage of time slots are available. On the other hand, the link quality determines the maximum throughput.  In addition, with a QoS guarantee, combining power and rate control to optimize the resource consumption is another research direction \cite{DongTWC2008} in cognitive radio networks.

%% file: conclusion.tex
\section{Conclusion}
\label{sec: conclusion}

In this paper, we provided a comprehensive survey of rate control mechanisms at the MAC-layer of 802.11 networks developed over the past two decades. While many solutions have been proposed to address different needs, only a handful of them have been adopted in practice. Among them, some had not been studied and evaluated systematically until recently. This survey analyzes existing solutions based on the two fundamental aspect in rate control --- metrics and algorithms, and shares insights that have impact on the development of new MAC-layer rate control mechanisms. Drawing observations from other published work, we further provided a comparison of the existing solutions and outlined their characteristics. We showed that the existing solutions for 802.11 networks can not be directly adopted in other types of systems and networks. The findings from this survey offer guidelines for the development of new rate control mechanisms for  emerging technologies and new applications. 

%% file: acknowledgement.tex
\section*{Acknowledgement}
The work is partially supported by the NSFC project 61702542. We would like to thank reviewers for their possitive comments.